\documentclass[preprint2]{aastex}

\shorttitle{Heating and cooling variations in M51}
\shortauthors{Parkin et al.}

\begin{document}


\title{Regional variations in the dense gas heating and cooling in M51 from \emph{Herschel}\footnote{\emph{Herschel} is an ESA space observatory with science instruments provided by European-led Principal Investigator consortia and with important participation from NASA.} far-infrared spectroscopy}


\author{T.~J. Parkin\altaffilmark{2}, C.~D. Wilson\altaffilmark{2}, M.~R.~P. Schirm\altaffilmark{2}, 
	M. Baes\altaffilmark{3}, M. Boquien\altaffilmark{4}, A. Boselli\altaffilmark{4},
	A. Cooray\altaffilmark{5,6}, D. Cormier\altaffilmark{7}, K. Foyle\altaffilmark{2},
	O.~\L. Karczewski\altaffilmark{8}, V. Lebouteiller\altaffilmark{9}, I. de~Looze\altaffilmark{3},
	S.~C. Madden\altaffilmark{9}, H. Roussel\altaffilmark{10}, M. Sauvage\altaffilmark{9},
	and L. Spinoglio\altaffilmark{11}}
\altaffiltext{2}{Department of Physics \& Astronomy, McMaster University, Hamilton, Ontario, L8S~4M1, Canada; \email{parkintj@mcmaster.ca}}
\altaffiltext{3}{Sterrenkundig Observatorium, Universiteit Gent, Krijgslaan 281 S9, B-9000 Gent, Belgium}
\altaffiltext{4}{Laboratoire d'Astrophysique de Marseille - LAM, Universit\'e d'Aix-Marseille \& CNRS, UMR7326, 38 rue F. Joliot-Curie, 13388 Marseille Cedex 13, France}
\altaffiltext{5}{Department of Physics and Astronomy, University of California, Irvine, CA 92697, USA}
\altaffiltext{6}{Division of Physics, Astronomy and Mathematics, California Institute of Technology, Pasadena, CA, 91125, USA}
\altaffiltext{7}{Institut f\"ur theoretische Astrophysik, Zentrum f\"ur Astronomie der Universit\"at Heidelberg, Albert-Ueberle Str. 2, D-69120 Heidelberg, Germany}
\altaffiltext{8}{Department of Physics and Astronomy, University College London,  
Gower Street, London WC1E 6BT, UK}
\altaffiltext{9}{CEA, Laboratoire AIM, Irfu/SAp, Orme des Merisiers, F-91191 Gif-sur-Yvette, France}
\altaffiltext{10}{Institut d'Astrophysique de Paris, UMR7095 CNRS, Universit\'e Pierre \& Marie Curie, 98 bis Boulevard Arago, F-75014 Paris, France}
\altaffiltext{11}{Istituto di Astrofisica e Planetologia Spaziali, INAF-IAPS, Via Fosso  
del Cavaliere 100, I-00133 Roma, Italy}




\begin{abstract}
We present \emph{Herschel} PACS and SPIRE spectroscopy of the most important far-infrared cooling lines in M51, [C\,\textsc{ii}](158~$\mu$m), [N\,\textsc{ii}](122 \& 205~$\mu$m), [O\,\textsc{i}](63 and 145~$\mu$m) and [O\,\textsc{iii}](88~$\mu$m). We compare the observed flux of these lines with the predicted flux from a photon dominated region model to determine characteristics of the cold gas such as density, temperature and the far-ultraviolet radiation field, $G_{0}$, resolving details on physical scales of roughly 600~pc.  We find an average [C\,\textsc{ii}]/$F_{\mathrm{TIR}}$ of $4 \times 10^{-3}$, in agreement with previous studies of other galaxies.  A pixel-by-pixel analysis of four distinct regions of M51 shows a radially decreasing trend in both the far-ultraviolet (FUV) radiation field, $G_{0}$ and the hydrogen density, $n$, peaking in the nucleus of the galaxy, then falling off out to the arm and interarm regions.  We see for the first time that the FUV flux and gas density are similar in the differing environments of the arm and interarm regions, suggesting that the inherent physical properties of the molecular clouds in both regions are essentially the same.
\end{abstract}


\keywords{galaxies:individual(M51) -- galaxies:ISM -- infrared:ISM -- ISM:lines and bands}



\section{Introduction}
The atomic fine structure lines [C\,\textsc{ii}](158~$\mu$m), [N\,\textsc{ii}](122 and 205~$\mu$m), [O\,\textsc{i}](63 and 145~$\mu$m) and [O\,\textsc{iii}](88~$\mu$m) are among the dominant cooling lines of the cold neutral and ionized regimes of the interstellar medium (ISM).  The progenitor atoms of these lines are collisionally excited then de-excite through forbidden transitions, emitting photons and thus removing thermal energy from the gas.  The [O\,\textsc{i}] lines originate in the neutral regime of photon dominated regions (PDRs) \citep{1985ApJ...291..722T} as atomic oxygen has an ionization potential greater than 13.6~eV, while [C\,\textsc{ii}] emission, though considered a primary tracer of PDRs, can also trace ionized gas as the ionization potential of atomic carbon is only 11.26~eV.  In contrast, the [N\,\textsc{ii}] and [O\,\textsc{iii}] lines only trace ionized gas, particularly in H\,\textsc{ii} regions, because the ionization potentials of N and O$^{+}$ are 14.5 and 35~eV, respectively, requiring the presence of a hard radiation field.  Thus, observations of these lines can tell us important characteristics about the gas in these components of the ISM.  

Observationally, these cooling lines were first detectable with the \emph{Kuiper Airborne Observatory} (KAO) and the \emph{Infrared Space Observatory} (ISO), and are now observable with the \emph{Herschel Space Observatory} \citep{2010A&A...518L...1P} at unprecedented resolution.  Previous surveys studied [C\,\textsc{ii}](158~$\mu$m), [N\,\textsc{ii}](122~$\mu$m), and [O\,\textsc{i}](63~$\mu$m) on global scales in galaxies using ISO and found a decreasing ratio of $L_{\mathrm{[CII]}}$/$L_{\mathrm{TIR}}$ with increasing infrared colour, $F_{\nu}(60 \mu\mathrm{m})/F_{\nu}(100 \mu\mathrm{m})$, as well as a correlation between $L_{\mathrm{[OI]63}}$/$L_{\mathrm{[C\,\textsc{ii}]}}$ and infrared colour \citep[e.g.][]{1997ApJ...491L..27M,2001ApJ...561..766M,2001A&A...375..566N,2008ApJS..178..280B}.  $L_{\mathrm{[CII]}}$/$L_{\mathrm{TIR}}$ is used as a probe of the photo-electric heating efficiency of the far-ultraviolet (FUV) radiation field, indicating the fraction of FUV photons that contributes to dust heating via absorption, versus the fraction responsible for ejecting electrons from dust grains or polycyclic aromatic hydrocarbons (PAHs) \citep{1985ApJ...291..722T, 2001ApJ...561..766M}, which thermally heat the gas.  The ratio $F_{\nu}(60 \mu\mathrm{m})/F_{\nu}(100 \mu\mathrm{m})$ indicates dust temperatures, with higher temperatures indicating compact H\,\textsc{ii} regions. These studies concluded the trends seen indicate a decrease in this heating efficiency with increasing color.

More recently these cooling lines have been probed on galactic and spatially resolved scales in nearby galaxies with \emph{Herschel} \citep[e.g.][]{2010A&A...518L..60B, 2012ApJ...751..144B, 2011ApJ...728L...7G, 2011A&A...532A.152M, 2012ApJ...747...81C, 2012A&A...544A..55B, 2012A&A...548A..91L, 2013A&A...549A.118C}.  \citet{2010A&A...518L..60B,2012ApJ...751..144B}, \citet{2011A&A...532A.152M} and \citet{2012A&A...548A..91L} investigated the Seyfert~1 galaxy NGC~1097, M33 and the H\,\textsc{ii} region LMC-N11, respectively, and found that $L_{\mathrm{[CII]}}$/$L_{\mathrm{TIR}}$ varies on local scales as well.  Furthermore, \citet{2012ApJ...747...81C} and \citet{2012A&A...548A..91L} also see the same relationship between $L_{\mathrm{[CII]+[OI]63}}$/$L_{\mathrm{TIR}}$ and infrared colour that is seen on global scales.  Interestingly, they also investigated the ratio $L_{\mathrm{[CII]+[OI]63}}$/$L_{\mathrm{PAH}}$ as a function of color, and found there was a stronger correlation between these two parameters than between $L_{\mathrm{[CII]+[OI]63}}$/$L_{\mathrm{TIR}}$ versus color, implying PAHs are likely a better indicator of heating efficiency than the total infrared luminosity within the gas.  At the warmest colors, $L_{\mathrm{[CII]+[OI]63}}$/$L_{\mathrm{PAH}}$ decreased slightly and the authors attributed this to the PAHs becoming increasingly ionized, thus reducing their ability to eject photoelectrons.

Diagnosing the physical conditions of the gas in the interstellar medium (ISM) using these important cooling lines requires comparing the observed fluxes to those predicted by a PDR model.  \citet{1985ApJ...291..722T} presented a PDR model that characterises the physical conditions in the PDR by two free variables, the hydrogen nucleus density, $n$, and the strength of the FUV radiation field in units of the Habing Field, $G_0 = 1.6 \times 10^{-3}$~erg~cm$^{-2}$~s$^{-1}$ \citep{1968BAN....19..421H}.  They assume a slab geometry and include a complex chemical network, thermal balance and radiative transfer.  In successive papers, \citet{1990ApJ...358..116W} and \cite{1991ApJ...377..192H} expanded this work to model ensembles of molecular clouds in diffuse environments as well as galactic nuclei and AGN.  \citet{1999ApJ...527..795K,2006ApJ...644..283K} have further updated the model of \citet{1990ApJ...358..116W} and provide a set of diagnostic plots using ratios between cooling lines and the total infrared luminosity to determine $n$ and $G_0$.\footnote{Also available for analysis using this model is the PDR Toolbox \citep{2008ASPC..394..654P}, found at http://dustem.astro.umd.edu/pdrt}  PDR models using different sets of observables have also been developed by \citet{1986ApJS...62..109V,1988ApJ...334..771V}, \citet{1989ApJ...338..197S, 1995ApJS...99..565S}, \citet{1997ApJ...482..298L}, \citet{2000A&A...358..682S} and \citet{2006ApJS..164..506L}.  PDR models with a spherical geometry instead of the plane-parallel geometry also exist, such as Kosma~$\tau$ \citep[e.g.][]{2006A&A...451..917R}.  For a recent discussion and comparison of the different available PDR models see \citet{2007A&A...467..187R}. A number of studies have compared observations to PDR models to determine the gas density, temperature and strength of the FUV radiation field in numerous galaxies and Galactic PDRs.  When they compared their observations to the PDR model of \citet{1999ApJ...527..795K}, \citet{2001ApJ...561..766M} found the FUV radiation field, $G_{0}$, scales as the hydrogen nucleus density, $n$, to the power 1.4, with $10^{2} \le G_{0} \le 10^{4.5}$ and $10^{2}\,\mathrm{cm}^{-3} \le n \le 10^{4.5}\,\mathrm{cm}^{-3}$ for their sample of galaxies.  \citet{2012ApJ...747...81C} studied NGC~1097 as well as NGC~4559, and found $10^{2.5}\,\mathrm{cm}^{-3} \le n \le 10^{3}\,\mathrm{cm}^{-3}$ across both galactic disks, with $50 \le G_{0} \le 1000$.

The goal of this paper is to investigate the gas component of the ISM in M51 with modelling of the far-infared fine-structure lines.  M51 \citep[$D = 9.9$~Mpc; ][]{2009AstL...35..599T} is a gas-rich, grand-design spiral galaxy with a smaller companion, NGC~5195, and is classified as a Seyfert~2 galaxy \citep{1997ApJS..112..315H}.   The metallicity (12+log(O/H)) of M51 is $8.55 \pm 0.01$ on average and has a slight radial gradient \citep{2010ApJS..190..233M}, though it does not change significantly over the area covered in this work.  M51 has been previously observed with the KAO and ISO to investigate the far-infrared cooling lines.  \citet{2001ApJ...561..203N} mapped the [C\,\textsc{ii}](158~$\mu$m) line in the inner $6 \arcmin$ of the galaxy with the far-infrared imaging Fabry-Perot interferometer on the KAO at $55 \arcsec$ resolution ($\sim$2.6~kpc at our adopted distance).  Their comparison with PDR models revealed $G_0 \sim 150-850$ and two density solutions, $n \sim 10^{2}$~cm$^{-3}$ and $n \sim 5 \times 10^{5}$~cm$^{-3}$. Later, \citet{2005A&A...441..961K} used ISO to map M51 in [C\,\textsc{ii}](158~$\mu$m), [O\,\textsc{i}](63~$\mu$m) and [N\,\textsc{ii}](122~$\mu$m).  This study focused on the nucleus and two selected positions in the spiral arms at a resolution of $80\arcsec$ ($\sim$3.8~kpc at our adopted distance).  Their comparison to PDR models revealed $G_{0}=20-30$ and $n \sim 10^{4}$~cm$^{-3}$ within these regions.  As part of the \emph{Herschel} Guaranteed Time Key Project the Very Nearby Galaxies Survey (VNGS; PI: C.~D. Wilson), \citet{2012ApJ...755..165M} presented a detailed analysis of the dust and gas of both M51 and NGC~5195 using \emph{Herschel} Photodetector Array Camera and Spectrometer \citep[PACS; ][]{2010A&A...518L...2P} and Spectral and Photometric Imaging Receiver \citep[SPIRE; ][]{2010A&A...518L...3G} photometry, as well as the spectral energy distribution (SED) model of \citet{2007ApJ...657..810D}.  They find that there was a burst of star formation approximately $370 - 480$~Myr ago, a gas-to-dust mass ratio of $94 \pm 17$ (the Milky Way has a gas-to-dust mass ratio of $\sim 160$ \citep{2004ApJS..152..211Z}), and an interstellar radiation field (ISRF) within the spiral arms of approximately 5 to 10 times the average value of the ambient ISRF in the Solar neighborhood ($G_{0} \sim 6-12$).

The general goals of the VNGS are to investigate properties of the gas and dust in the ISM in an intentionally diverse sample of 13 nearby galaxies using \emph{Herschel}.  The galaxies represent a sample of different morphological types and have previously been observed in numerous other wavebands across the electromagnetic spectrum, thus allowing us to create a complete picture of the conditions in the ISM of these objects.  Here we present new far-infrared spectroscopy of M51 from the PACS instrument, focusing on the [C\,\textsc{ii}](158~$\mu$m), [N\,\textsc{ii}](122~$\mu$m), [O\,\textsc{i}](63 and 145~$\mu$m) and [O\,\textsc{iii}](88~$\mu$m) fine structure lines at unprecedented resolution (better than $\sim 12\arcsec$, or roughly 600~pc).  In addition we present observations of the [N\,\textsc{ii}](205~$\mu$m) line from the SPIRE Fourier Transform Spectrometer (FTS).  We use these spectra to investigate the gas component of the galaxy by using the PDR model of \citet{1999ApJ...527..795K, 2006ApJ...644..283K} to diagnose some of the physical characteristics of the ISM.  In Section~\ref{Herschel_obs} we describe our method for processing the data.  In Section~\ref{gas_char} we describe the characteristics of the gas and in Section~\ref{pdr_model} we compare our observations to theoretical PDR models.  We conclude in Section~\ref{conclusions}.

\section{\emph{Herschel} Observations}\label{Herschel_obs}

\subsection{PACS observations}\label{pacs_obs}
The PACS spectrometer covers a wavelength range of 51 to 220~$\mu$m, and comprises 25 spatial pixels (spaxels) arranged in a $5 \times 5$ grid with a square field of view on the sky of 47$\arcsec$ on a side.  Each spaxel records a separate spectrum from a 9.4$\arcsec$ field of view at a spectral resolution ranging from approximately 75 to 300~km~$s^{-1}$.\footnote{PACS Observer's Manual (hereafter PACS~OM; HERSCHEL-HSC-DOC-0832, 2011), available for download from the ESA \emph{Herschel} Science Centre.}  The FWHM of the beam varies from just over 9$\arcsec$ to about 13$\arcsec$.  More details about the spectrometer can be found in \citet{2010A&A...518L...2P} or in the PACS~OM.

All of our PACS spectroscopic observations of M51 were carried out as part of the VNGS using the unchopped grating scan mode.  We have raster maps covering the central $2.5\arcmin \times 2.5\arcmin$ of M51 in the [C\,\textsc{ii}](158 $\mu$m), [N\,\textsc{ii}](122 $\mu$m) and [O\,\textsc{i}](63 $\mu$m) lines (hereafter [C\,\textsc{ii}], [N\,\textsc{ii}]122 and [O\,\textsc{i}]63, respectively), and $47\arcsec \times 3.25\arcmin$ raster strips extending from the centre (for the [O\,\textsc{i}](145 $\mu$m) and [O\,\textsc{iii}](88 $\mu$m) lines (hereafter [O\,\textsc{i}]145 and [O\,\textsc{iii}], respectively)) and $47\arcsec \times 2.25\arcmin$ raster strips extending from the near centre (for the [C\,\textsc{ii}], [N\,\textsc{ii}]122 and [O\,\textsc{i}]63 lines) of the galaxy out along a position angle of 310 degrees counter-clockwise from north.   In Figure~\ref{fig:Ltir} we show the outline of our spectroscopic maps on the total infrared flux map (see Section~\ref{ancillary} for details).  For the central maps we use the recommended raster point step size of 24$\arcsec$ (16$\arcsec$) and raster line step size of 22$\arcsec$ (14.5$\arcsec$) for full Nyquist sampling of the red (blue) wavebands (PACS~OM).  The strips have 30$\arcsec$ raster spacing along the orientation angle.  The basic observational details are summarized in Table~\ref{table: herschel_char}.

The PACS spectroscopic observations were processed using the Herschel Interactive Processing Environment \citep[HIPE; ][]{2010ASPC..434..139O} developer's track 9.0 build 2634 using calibration version FM, 32.  We follow the standard pipeline reduction steps for the unchopped observing mode from Level 0 to Level 1. The data were flagged again, and the unbinned spectral data fit with a second order polynomial for the continuum and a Gaussian function for the spectral line using the PACSman package \citep{2012A&A...548A..91L}.  Lastly, we combine the individual rasters to produce a final integrated flux mosaic map by projecting each raster onto an oversampled grid, also using PACSman.

\begin{deluxetable}{lcccccc}
\tabletypesize{\small}
\tablecolumns{7}
\tablecaption{Properties of our \emph{Herschel} observations\label{table: herschel_char}}
\tablewidth{0pt}
\tablehead{
\colhead{Line} & \colhead{Wavelength} & \colhead{OBSID} & \colhead{Date of} & \colhead{Map Size}
 	& \colhead{FWHM\tablenotemark{a}} & \colhead{Integration} \\
                   & \colhead{($\mu$m)}   &            & \colhead{Observation}
    & \colhead{$\arcmin\times\arcmin$} & \colhead{($\arcsec$)} & \colhead{Time (s)}}
 \startdata
 [O\,\textsc{i}]              & 63.184     & 1342211190 & 2010 Dec 14 & 2.5~$\times$~2.5
 	& $\sim$9.3             & 10735 \\
 		           & 		    & 1342211195 & 2010 Dec 15  & 0.72~$\times$~2.25
 	& $\sim$9.3             & 1254 \\
 $[$O\,\textsc{iii}$]$ & 88.356     & 1342211191 & 2010 Dec 14 & 0.72~$\times$~3.25
 	& $\sim$9.3             & 2791 \\
 $[$N\,\textsc{ii}$]$  & 121.898    & 1342211189 & 2010 Dec 14 & 2.5~$\times$~2.5
 	& $\sim$10              & 10511  \\
                   &            & 1342211192 & 2010 Dec 15 & 0.72~$\times$~2.25
    & $\sim$10              & 2005  \\
 $[$O\,\textsc{i}$]$   & 145.525    & 1342211194 & 2010 Dec 15 & 0.72~$\times$~3.25
 	& $\sim$11              & 5277  \\
 $[$C\,\textsc{ii}$]$  & 157.741    & 1342211188 & 2010 Dec 14 & 2.5~$\times$~2.5
 	& $\sim$11.5            & 5597  \\
                   &            & 1342211193 & 2010 Dec 15 & 0.72~$\times$~2.25
    & $\sim$11.5            & 1254 \\
 $[$N\,\textsc{ii}$]$  & 205.178    & 1342201202 & 2010 Jul 25 & $\sim2\arcmin$~diameter circle
 	& $17$                  & 17603 \\
 \enddata
 \tablenotetext{a}{Values are from the PACS Observer's Manual and the SPIRE Observers' Manual.}
\end{deluxetable}

\subsection{SPIRE observations}\label{spire_obs}
As characterized in \citet{2010A&A...518L...3G} and the SPIRE Observers' Manual\footnote{Hereafter SPIRE~OM.  Document HERSCHEL-DOC-0798 version 2.4 (June 2011), is available from the ESA \emph{Herschel} Science Centre}, the \emph{Herschel} SPIRE FTS instrument consists of two bolometer arrays, the SPIRE Short Wavelength spectrometer array (SSW) covering the wavelength range 194 to 313~$\mu$m, and the SPIRE Long Wavelength spectrometer array (SLW) covering the range 303 to 671~$\mu$m, and has a field of view approximately 2$\arcmin$ in diameter.  Each array is arranged in a honeycomb pattern, with 37 (19) receivers in the SSW (SLW) with receiver beams spaced by 33$\arcsec$ (51$\arcsec$) on the sky, thus sampling different regions of the overall field of view. The spectral resolution ranges from approximately 1.2~GHz (highest attainable resolution) to upwards of 25~GHz, and is dependent on the maximum difference between path lengths, $d$, travelled by the radiation after it has passed through the interferometer's beam splitter, as given by $(2d)^{-1}$.  The FWHM of the beam varies as a function of wavelength between 17 and 21$\arcsec$ (29 and 42$\arcsec$) for the SSW (SLW) \citep{2010A&A...518L...3G,2010A&A...518L...4S}.

The \emph{Herschel} FTS observation of the [N\,\textsc{ii}](205~$\mu$m) line (hereafter [N\,\textsc{ii}]205) is also part of the VNGS and its basic properties are listed in Table~\ref{table: herschel_char}.  The observations consist of a single pointing carried out with intermediate spatial sampling covering a circular area approximately 2$\arcmin$ in diameter, and the high spectral resolution (1.2~GHz) setting. The total integration time of the observation is $\sim$5 hours, with 32 repetitions of the spectral scan pair at each of the four jiggle positions that produce an intermediately sampled map.  The standard processing pipeline for intermediately sampled mapped observations using HIPE 9.0 and SPIRE calibration context v9.1 was used, starting with the Level~0.5 product.   Basic standard processing steps include first level deglitching, and correcting the signal for the non-linearity of detector response, temperature drifts, saturation effects and time domain phase delays.  Next, Level~1 interferograms are created, and the baseline is subsequently fit and subtracted from the signal.  The interferograms then undergo second level deglitching, phase corrections, and a Fourier Transform to produce the spectra.  The spectra are then calibrated to convert units to flux, telescope and instrument emission is removed, and an extended source flux conversion is applied.  One additional step is applied to our data at this stage that is not part of the standard pipeline for mapped observations.  A point source flux calibration correction that is determined separately for each bolometer of the array is applied to the data because M51 does not uniformly fill the beam.  Finally, two (one each for the SSW and SLW) spectral cubes containing the processed spectra are created.

The final spectra were fit with a polynomial and Sinc function for the baseline and line, respectively (see Schirm et al. 2013, in prep for details). Given that the spatial sampling is intermediate, we created the final integrated flux map using a fine, $4\arcsec$, pixel scale such that the finite pixels are centred on each of the bolometers of the FTS array, while the remaining pixels are left blank.  The calibration uncertainty for the [N\,\textsc{ii}]205 map is better than 7\%, and stems from a comparison between a model spectrum of Uranus and observational and model spectra of Neptune, pointing uncertainties, and the accuracy of background signal removal (SPIRE~OM).

\subsection{Ancillary Data}\label{ancillary}
\begin{figure}[h]
\includegraphics[width=\columnwidth]{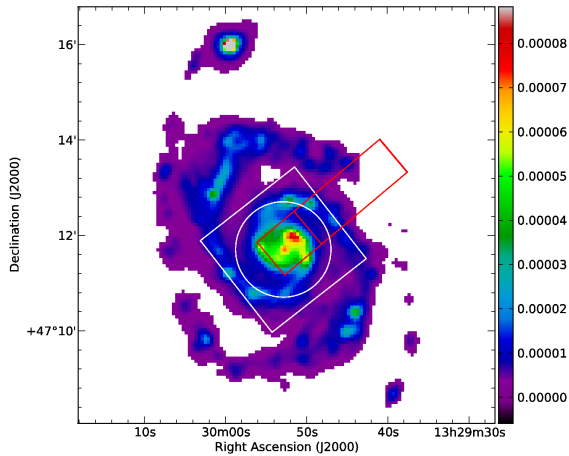}
\caption{The total infrared flux, $F_{\mathrm{TIR}}$, calculated using the MIPS~24~$\mu$m, PACS~70 and 160~$\mu$m maps, and Equation~\ref{eqn:Ftir}.  The map is at the resolution of the PACS 160~$\mu$m map, 12$\arcsec$ and has a plate scale of 4$\arcsec$.  Units are W~m$^{-2}$~sr$^{-1}$.  The white square and red rectangles outline the coverage of the PACS spectroscopic maps and strips, respectively.  The white circle represents the footprint of the SPIRE FTS observations.}
\label{fig:Ltir}
\end{figure}

We also make use of the PACS photometric maps at 70 and 160~$\mu$m originally presented in \citet{2012ApJ...755..165M}, as well as the MIPS 24~$\mu$m map from the \emph{Spitzer Space Telescope}, which was re-processed by \citet{2012MNRAS.423..197B} for the purposes of complementing the \emph{Herschel} photometry in a couple of Guaranteed Time programs, including the VNGS.  We calculate the total infrared flux using the MIPS 24~$\mu$m, PACS 70 and 160~$\mu$m maps, and the empirically determined equation for the total infrared flux (or luminosity) from \citet{2002ApJ...576..159D},
\begin{eqnarray}\label{eqn:Ftir}
F_{\mathrm{TIR}} = \xi_{1}\nu F_{\nu}(24 \mu\mathrm{m}) \, + \, \xi_{2}\nu F_{\nu}(70 \mu\mathrm{m})
	\nonumber \\
	+ \, \xi_{3}\nu F_{\nu}(160 \mu\mathrm{m}),
\end{eqnarray}
where [$\xi_{1}$, $\xi_{2}$, $\xi_{3}$] = [1.559, 0.7686, 1.347] at a redshift of $z=0$.  The total infrared luminosity determined with this equation covers emission from 3 to 1100$\mu$m.  The map of the total infrared flux, $F_{\mathrm{TIR}}$, is presented in Figure~\ref{fig:Ltir}, which also shows the spatial coverage of our PACS and SPIRE spectroscopic maps.

The \emph{Spitzer} IRAC 8~$\mu$m map of M51 was also obtained from the SINGS survey \citep{2003PASP..115..928K} archive to be used as a proxy for PAH emission.  We applied a color correction to the image following the method described in the \emph{Spitzer} Data Analysis Cookbook\footnote{Available for download at http://irsa.ipac.caltech.edu/data/SPITZER/\-docs/dataanalysistools/cookbook/.}, and subtracted the stellar contribution to the map using the correction from \citet{2010ApJ...715..506M}, which is adopted by \citet{2012ApJ...747...81C} to be an estimate of the total PAH power when only the IRAC 8$\mu$m map is available (their equation (2)).

\subsection{Data treatment for analysis}\label{data_treatment}
All of our spectral maps have been convolved with a Gaussian function to a common resolution matching that of the PACS 160~$\mu$m map, 12$\arcsec$.  The MIPS 24 and PACS 70~$\mu$m maps were convolved with the appropriate kernels developed by \citet{2011PASP..123.1218A}.  In addition, each map has been regridded to match the pixel size of the PACS 160~$\mu$m map, 4$\arcsec$.  For the comparison with the PDR models, we applied a 5$\sigma$ cut to our PACS spectroscopic maps to ensure we are considering robust detections in our ratio maps.  Our quoted uncertainties throughout the paper take into account both measurement uncertainties in calculating the line fluxes for our PACS spectroscopic maps, and calibration uncertainties, unless otherwise noted.  We note here that the calibration uncertainties for the PACS spectroscopy are approximately 30\% and are dominated by small offsets in pointing as well as drifting of the detector response, while the calibration uncertainties for the photometry at 70 and 160~$\mu$m are 3 and 5\%, respectively (PACS OM).


\section{Physical characteristics of the gas}\label{gas_char}
\subsection{Line Emission Morphology}\label{morphology}
\begin{figure*}
 \includegraphics[width=7.0cm]{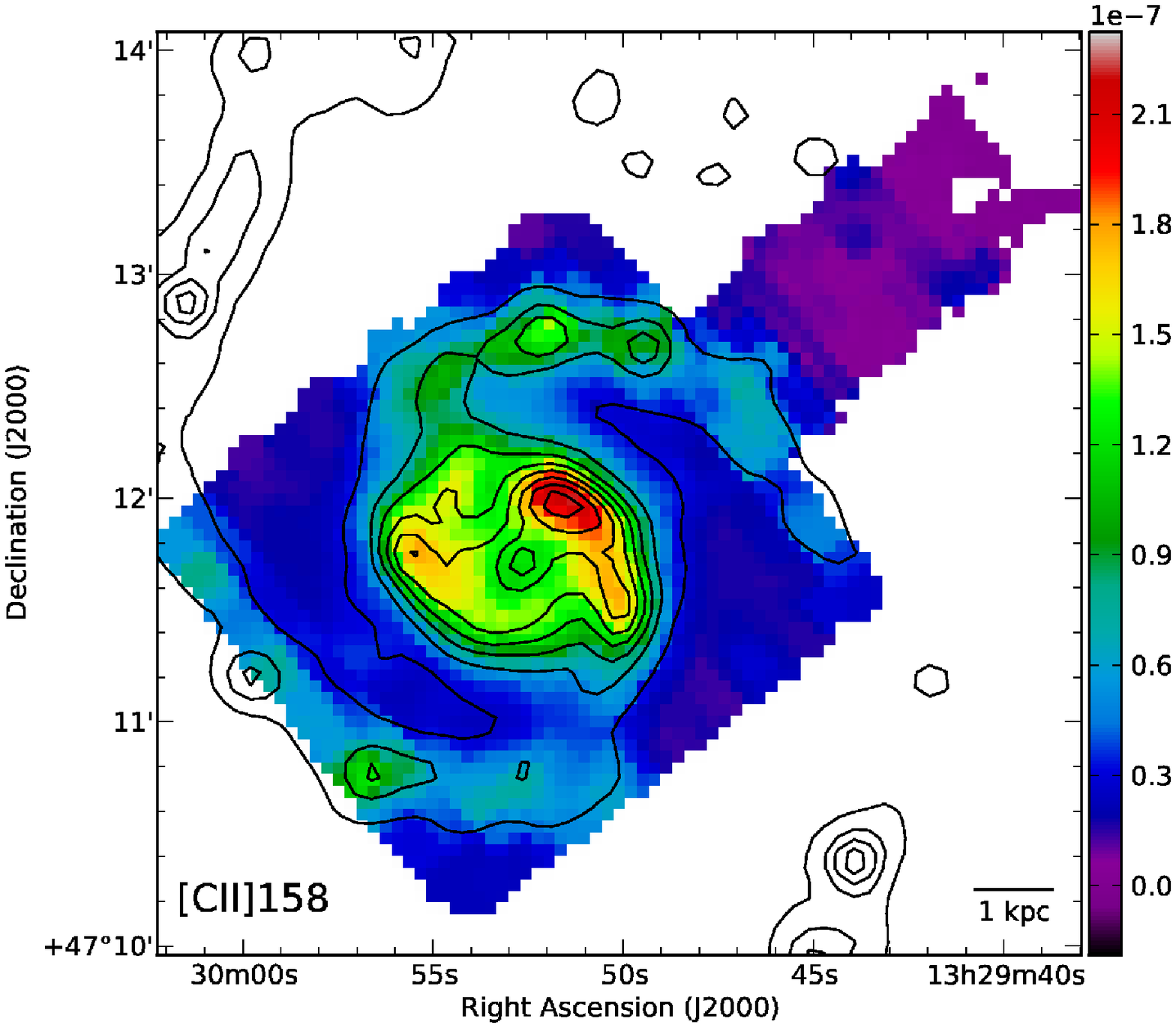}
 \includegraphics[width=7.0cm]{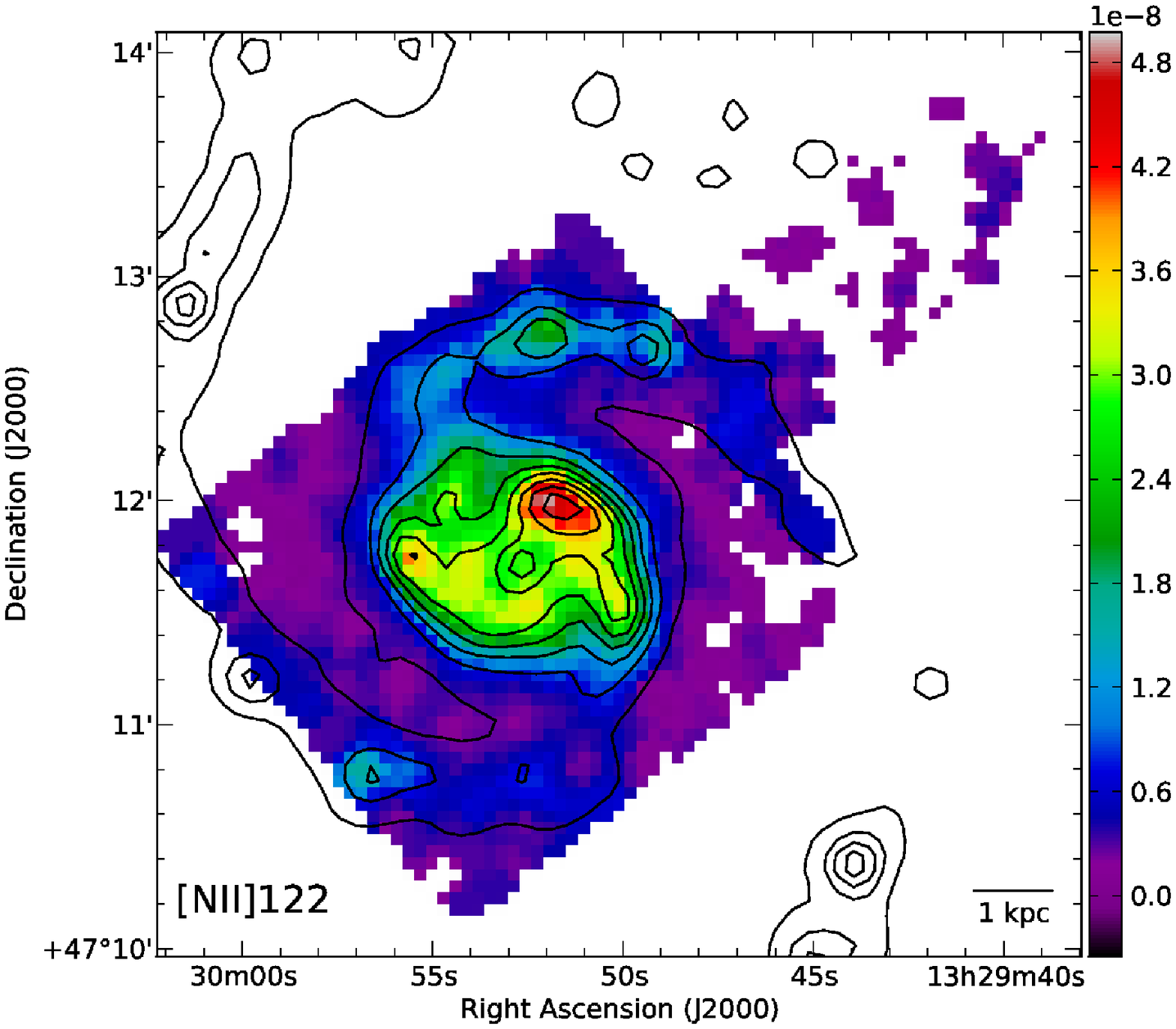}
 \includegraphics[width=7.0cm]{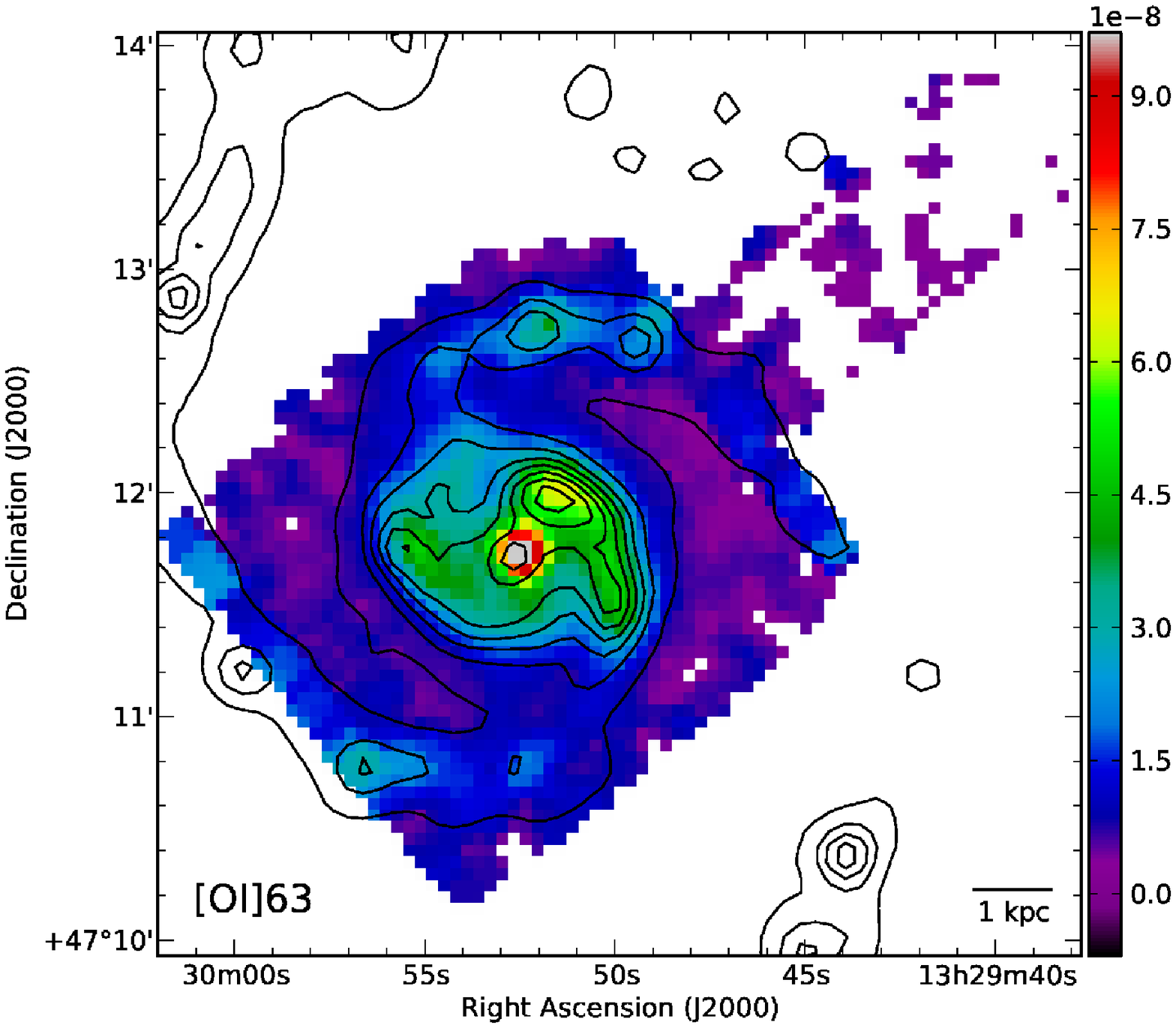}
 \includegraphics[width=7.0cm]{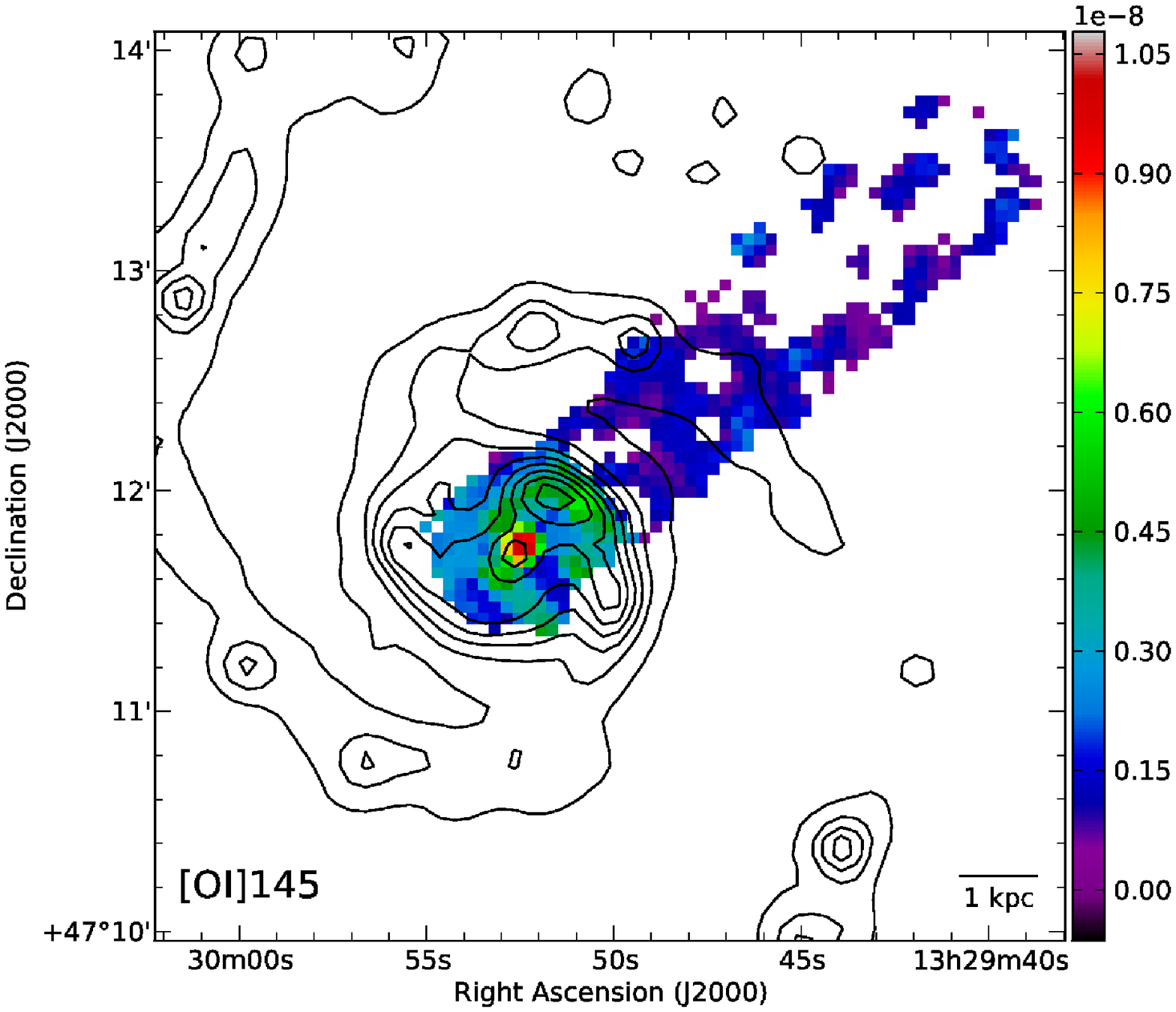}
 \includegraphics[width=7.0cm]{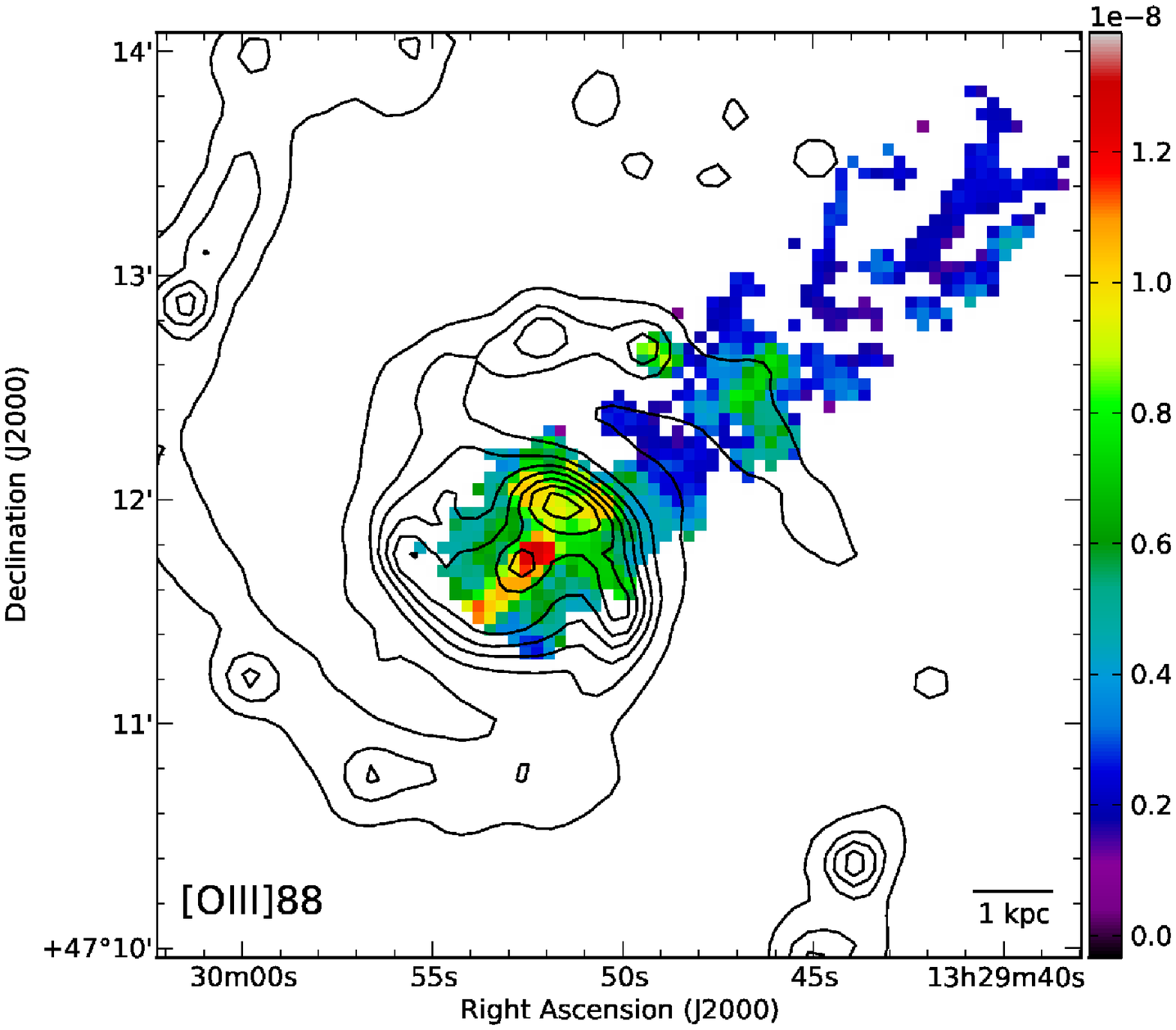}
 \includegraphics[width=7.0cm]{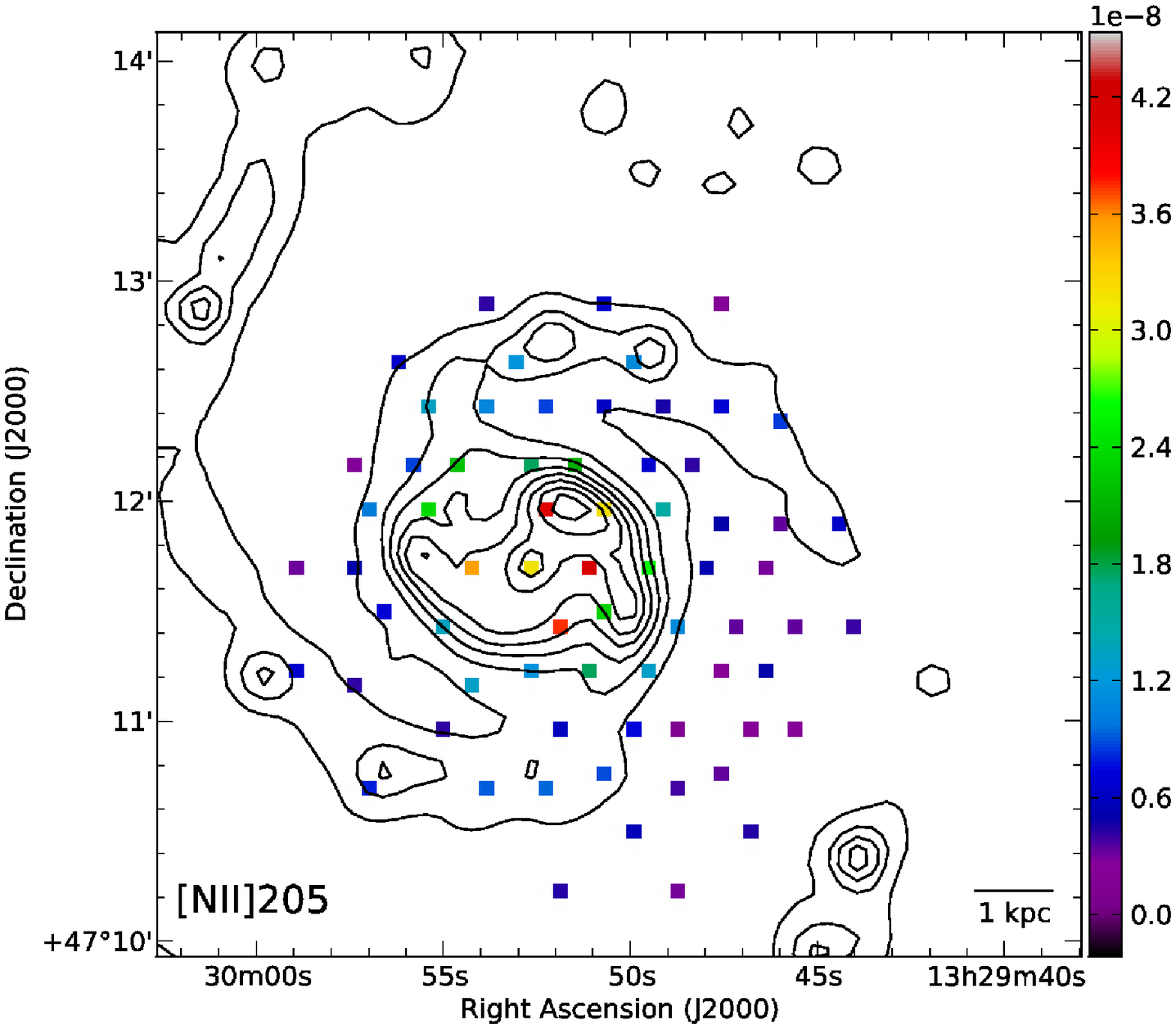}
\caption{\emph{Herschel} PACS and SPIRE spectroscopic maps of M51 of the six fine-structure lines at their native resolution and pixel scale.  We have applied a 3$\sigma$ cutoff to all five PACS images and the units are W~m$^{-2}$~sr$^{-1}$.  Contours of the total infrared flux are overlaid to direct the eye to the major features of the inner galaxy.  We note that the integrated intensity within each pixel of the [N\,\textsc{ii}]205 map is actually the average surface brightness over the 17$\arcsec$ beam of each bolometer in the FTS array, and not the average over the 4$\arcsec$ pixel each bolometer is centered on.}
\label{fig:pacs_spec_maps}
\end{figure*}

We present the final maps of the [C\,\textsc{ii}], [N\,\textsc{ii}]122, [O\,\textsc{i}]63, [O\,\textsc{i}]145, [O\,\textsc{iii}] and [N\,\textsc{ii}]205 lines at their native resolution with a 3$\sigma$ cut applied to the PACS maps in Figure~\ref{fig:pacs_spec_maps}.  We note that the [C\,\textsc{ii}] data were first presented by \citet{2013arXiv1304.1801S} using a different processing method, though it does not play a major role in their analysis. Both the [C\,\textsc{ii}] and [N\,\textsc{ii}]122 maps show similar distributions throughout the center of the galaxy.  Overall there is strong emission in the central $\sim1.25\arcmin$, with the innermost sections of the spiral arms showing the strongest emission.  The average signal-to-noise in the central region is $\sim200$ and 70 for the [C\,\textsc{ii}] and [N\,\textsc{ii}]122 maps, respectively.  The peak in the inner northwestern arm is also present in other wavebands including the 24, 70 and 160~$\mu$m images (combined in Figure~\ref{fig:Ltir}), as well as other star formation tracers such as H$\alpha$ and Pa$\alpha$ \citep[e.g.][]{2003PASP..115..928K,2005ApJ...633..871C}.   The emission is a factor of $\sim$1.5 higher in this peak compared to the center of the galaxy in both [C\,\textsc{ii}] and [N\,\textsc{ii}]122.  The spiral arms can be seen extending outward in weaker emission with a few stronger pockets.

The total [C\,\textsc{ii}] emission in our map is $(2.9 \pm 0.9) \times 10^{-14}$~W~m$^{-2}$ covering an area of $\sim5.5 \times 10^{-7}$~sr$^{-1}$.  \citet{2001ApJ...561..203N} mapped the [C\,\textsc{ii}] emission over a 6$\arcmin \times 6\arcmin$ area of M51 with the KAO and found a peak intensity of $(1.31 \pm 0.15) \times 10^{-4}$~ergs~s$^{-1}$~cm$^{-2}$~sr$^{-1}$ in the center of the galaxy.  The integrated intensity of our map over an aperture of approximately 1$\arcmin$ centered on the galaxy (to match the 55$\arcsec$ beam of the KAO) is $\sim (1.4 \pm 0.4) \times 10^{-4}$~ergs~s$^{-1}$~cm$^{-2}$~sr$^{-1}$, in good agreement with the KAO measurement.  \citet{2005A&A...441..961K} and \citet{2001A&A...375..566N} both present ISO observations of M51 and found a [C\,\textsc{ii}] integrated intensity at the center of $4.41 \times 10^{-5}$~ergs~s$^{-1}$~cm$^{-2}$~sr$^{-1}$ and $(9 \pm 2) \times 10^{-5}$~ergs~s$^{-1}$~cm$^{-2}$~sr$^{-1}$ within an 80$\arcsec$ beam, respectively, while in an aperture of the same size we measure an integrated intensity of $(1.1 \pm 0.3) \times 10^{-4}$~ergs~s$^{-1}$~cm$^{-2}$~sr$^{-1}$.  Our results agree with those of \citet{2001A&A...375..566N} but are higher than those of \citet{2005A&A...441..961K} by about a factor of two.  This discrepancy arises because \citet{2005A&A...441..961K} applied an extended source correction to their observations to obtain the integrated intensity within the ISO beam.

The [O\,\textsc{i}]63 map shows a strong spiral arm morphology in the innermost region (average signal-to-noise $\sim$40), with a similar peak in the northwest arm to that seen in [C\,\textsc{ii}] and [N\,\textsc{ii}]122.  However, in contrast to the [C\,\textsc{ii}] and [N\,\textsc{ii}]122 emission, the [O\,\textsc{i}]63 emission peaks prominently in the center of the galaxy.  M51 has been classified as a Seyfert~2 \citep{1997ApJS..112..315H}, and thus we believe the peaked emission in the center may be due to a low-luminosity active nucleus.  We measure a total [O\,\textsc{i}]63 flux within an 80$\arcsec$ beam of $(3 \pm 1) \times 10^{-15}$~W~m$^{-2}$, in good agreement with the values obtained with ISO of $(4.4 \pm 0.9) \times 10^{-15}$~W~m$^{-2}$ \citep{2001A&A...375..566N} and $\sim3.9 \times 10^{-15}$~W~m$^{-2}$ \citep{2005A&A...441..961K}.  \citet{2001A&A...375..566N} presented a survey of 34 galaxies including those classified as AGN, or Seyfert types.  Comparison of our [O\,\textsc{i}]63 flux obtained for M51 with values from their survey for AGN and Seyfert galaxies shows agreement within a factor of $\sim2$ for most sources, suggesting our results are typical for galaxies with active centers.

As we have only observed M51 along a strip in [O\,\textsc{i}]145 and [O\,\textsc{iii}] emission, the maps cover much less area.  [O\,\textsc{iii}] emission is concentrated primarily in the nuclear region with a peak in the very center, and a slight enhancement in the northwestern inner region of the spiral arm.  There is also a ``stripe'' extending from the center to the southeast that may indicate a bar-like structure.  Overall, the flux in this line is weaker than in [C\,\textsc{ii}], [N\,\textsc{ii}]122 and [O\,\textsc{i}]63 emission, in part due to the fact that the line is intrinsically weak.  However, there is also less sensitivity in the strips than the maps due to the raster spacing in the strips.  The average signal-to-noise in the central part of the [C\,\textsc{ii}] map is $\sim$200, while it is only about 8 and 11 for the central footprint in the [O\,\textsc{iii}] and [O\,\textsc{i}]145 strips, respectively.  The morphology of the center of the [O\,\textsc{i}]145 strip looks very similar to the [O\,\textsc{i}]63 emission, only weaker.  We note here that \citet{2001A&A...375..566N} observed the [O\,\textsc{i}]145 line with ISO but did not detect it, while they measured a flux of $(0.8 \pm 0.3) \times 10^{-15}$~W~m$^{-2}$ for the [O\,\textsc{iii}] line, twice as high as our measured value of $(3 \pm 1) \times 10^{-16}$~W~m$^{-2}$.  This is likely because we only observed a strip in this line and thus the 80$\arcsec$ aperture of \citet{2001A&A...375..566N} is larger than our observed region.

In Table~\ref{table:total_flux} we compare the total flux of each line with previous work.  In a region approximately $2.\arcmin7$ from the center of M51, centered on the H\textsc{ii} region CCM~10, \citet{2004AJ....128.2772G} used ISO to observe the same fine-structure lines as we present here.  \citet{2005A&A...441..961K} also observed M51 with ISO in the center and two locations in the spiral arms, detecting [C\,\textsc{ii}], [O\,\textsc{i}]63 and [N\,\textsc{ii}]122, while \citet{2001A&A...375..566N} presented observations of the center taken with ISO and detected the same lines we have, with the exception of the [O\,\textsc{i}]145 line.  To compare our results to these previous observations we have calculated the flux within the central 80$\arcsec$ for each line to match the beam size of a single pointing with ISO.  In general our measurements agree well with those of \citet{2001A&A...375..566N}; however, our values are stronger than those of \citet{2004AJ....128.2772G}.  This is a reasonable result given that CCM~10 is located in a region outside the central 80$\arcsec$ aperture and in fact falls outside our observed region entirely.

\begin{deluxetable}{lcccc}
\tabletypesize{\small}
\tablecolumns{5}
\tablecaption{Comparison to previous measurements of M51\label{table:total_flux}}
\tablewidth{0pt}
\tablehead{
\colhead{Line} & \multicolumn{4}{c}{Flux ($10^{-16}$~W~m$^{-2}$)} \\
\colhead{}     & \colhead{This work} & \colhead{\citet{2004AJ....128.2772G}}
		& \colhead{\citet{2001A&A...375..566N}} & \colhead{\citet{2005A&A...441..961K}} \\
\colhead{}	   & \colhead{central 80$\arcsec$\tablenotemark{a}} & \colhead{CCM~10\tablenotemark{b}}
		       & \colhead{center} & \colhead{center\tablenotemark{c}}}
 \startdata
 $[$C\,\textsc{ii}$]$(158~$\mu$m) & $135 \pm 40$  & $39 \pm 3$    & $104 \pm 21$ & $52.92$ \\
 $[$N\,\textsc{ii}$]$(122~$\mu$m) & $24 \pm 7$    & $3.3 \pm 0.1$ & $21 \pm 4$   & 14.76   \\
 $[$O\,\textsc{i}$]$(63~$\mu$m)   & $30 \pm 10$   & $14 \pm 4$    & $44 \pm 9$   & 38.64   \\
 $[$O\,\textsc{i}$]$(145~$\mu$m)  & $1.5 \pm 0.4$ & $0.8 \pm 0.2$ & \nodata      & \nodata \\
 $[$O\,\textsc{iii}$]$(88~$\mu$m) & $3 \pm 1$     & $8 \pm 2$     & $8 \pm 3$    & \nodata \\
 \enddata
 \tablecomments{Total flux measured within an 80$\arcsec$ aperture centered on the nucleus of M51.  The aperture matches the beam size of the ISO observations.  Only pixels with a 5$\sigma$ detection or better within the aperture were included in calculating the total flux from our \emph{Herschel} maps.}
 \tablenotetext{a}{The number of pixels with at least a 5$\sigma$ detection within the 80$\arcsec$ aperture varies between lines.  The [C\,\textsc{ii}], [N\,\textsc{ii}] and [O\,\textsc{i}]63 lines are detected in all pixels within the aperture, covering a total solid angle of $1.2 \times 10^{-7}$~sr.  The solid angle covered by the [O\,\textsc{i}]145 emission within the aperture is $4.6 \times 10^{-8}$~sr and the solid angle covered by the [O\,\textsc{iii}] emission within the aperture is $4.9 \times 10^{-8}$~sr.}
 \tablenotetext{b}{CCM~10 is an H\,\textsc{ii} region that lies outside the region discussed in this work.}
 \tablenotetext{c}{This is the same data as presented by \citet{2001A&A...375..566N}; however, an extended source correction has been applied by \citet{2005A&A...441..961K} in calculating the integrated intensity, reducing its value.}
\end{deluxetable}

\subsection{Line deficits}\label{subsec:ratio}
\begin{figure}
\includegraphics[width=7.0cm]{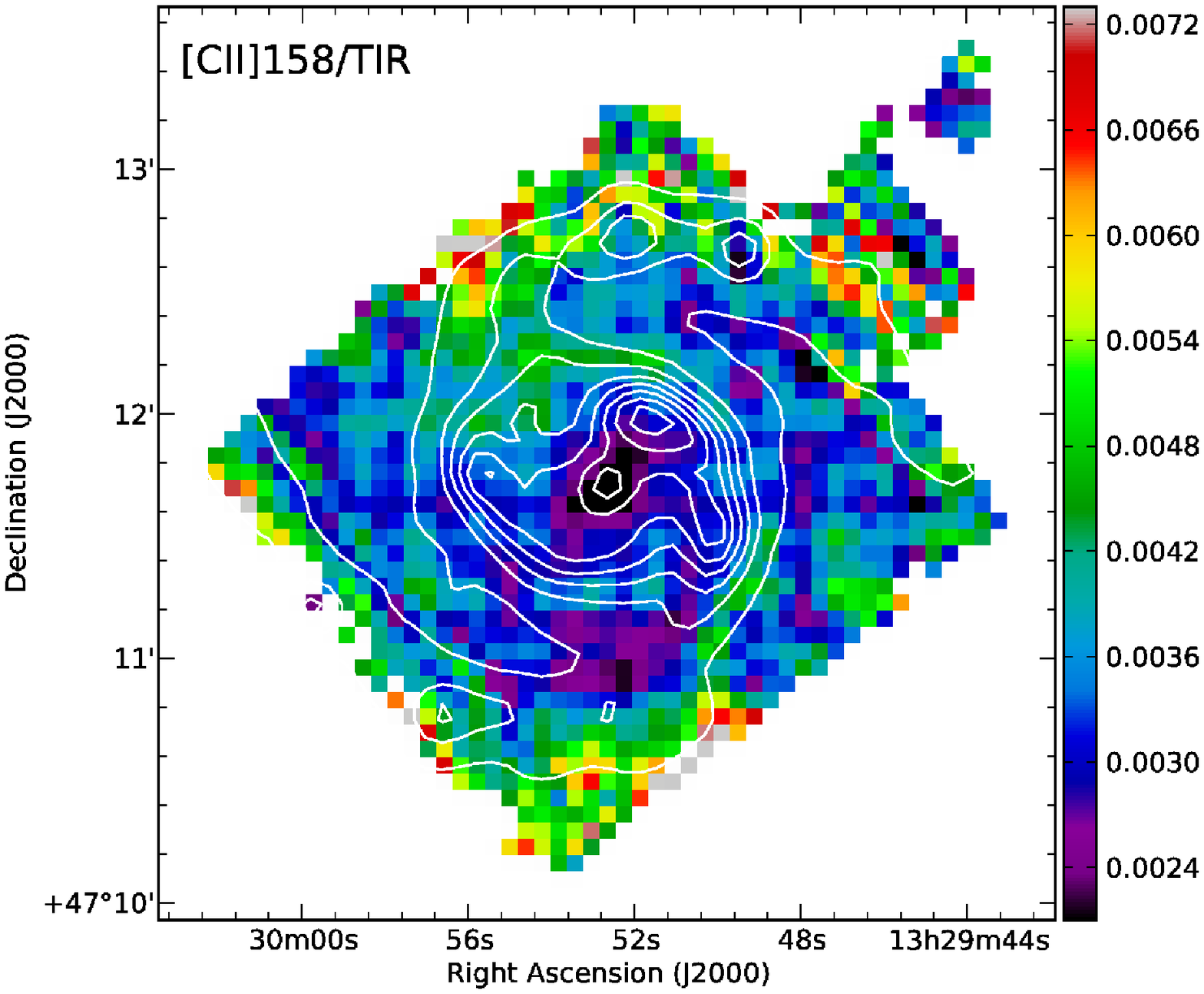}
\includegraphics[width=7.0cm]{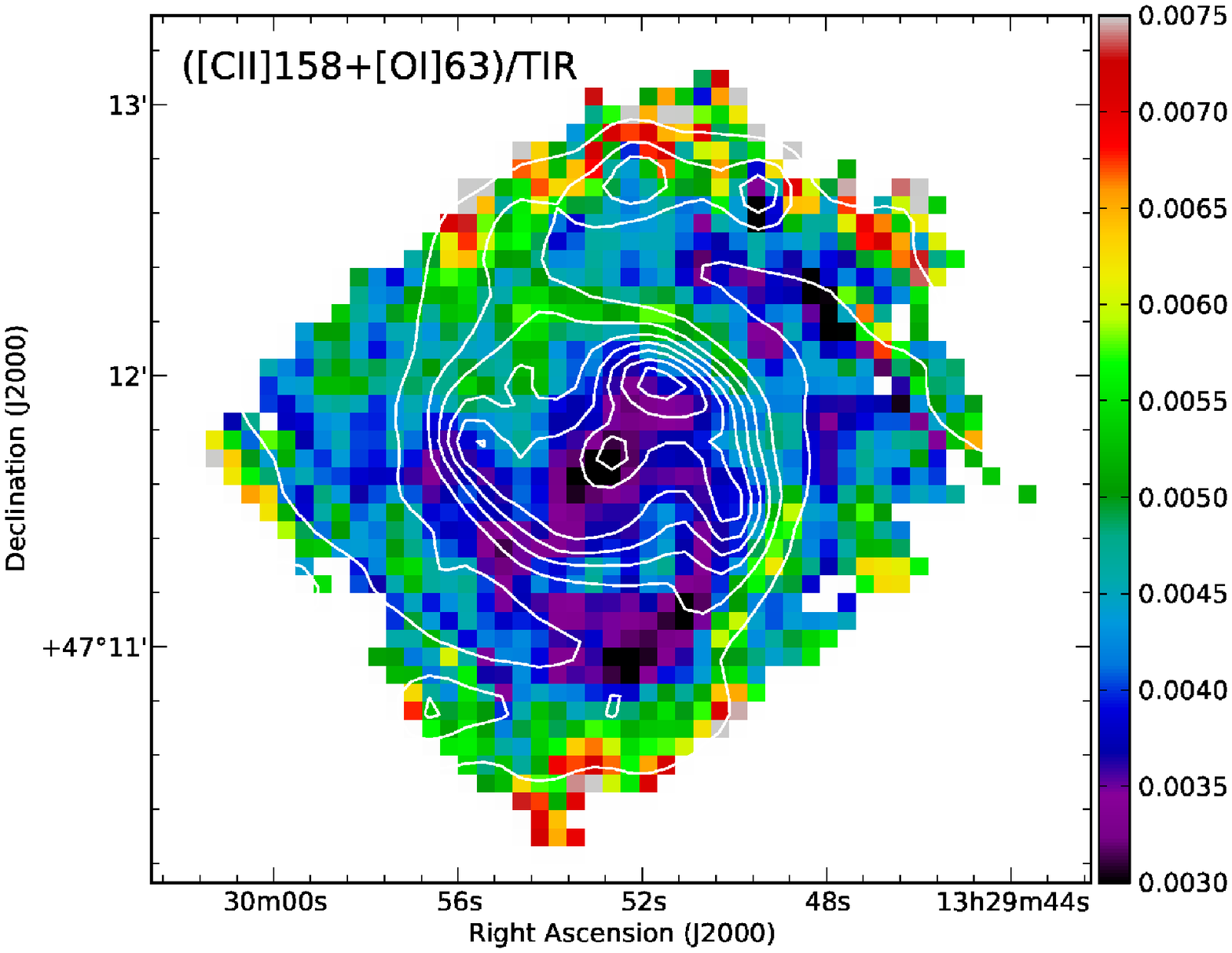}
\includegraphics[width=7.0cm]{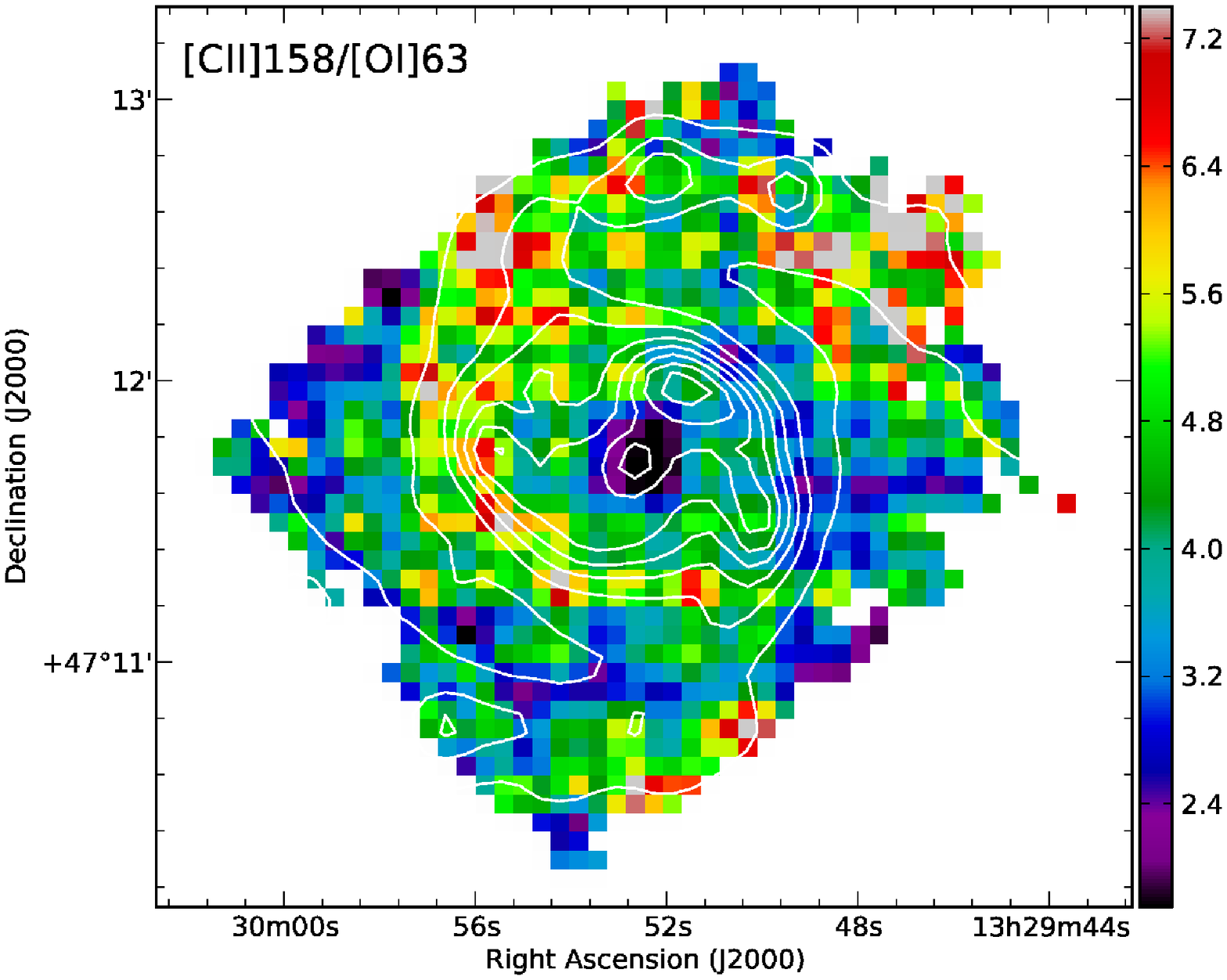}
\caption{\emph{top:} The [C\,\textsc{ii}] emission divided by the total infrared flux, $F_{\mathrm{TIR}}$, in M51.  \emph{middle: }The sum of the [C\,\textsc{ii}] and [OI]63 emission in M51 divided by the total infrared flux, $F_{\mathrm{TIR}}$.  \emph{bottom:} The [C\,\textsc{ii}] emission in M51 divided by the [O\,\textsc{i}]63 emission.  A 5$\sigma$ cutoff has been applied to all of the spectral line maps.  Contours of $F_{\mathrm{TIR}}$ are overlaid on each ratio plot to highlight the major features of the galaxy.}
\label{fig:ratio_plots}
\end{figure}

In Figure~\ref{fig:ratio_plots} (top) we show the [C\,\textsc{ii}]/$F_{\mathrm{TIR}}$ ratio, which varies between $10 \times 10^{-4}$ and $100 \times 10^{-4}$ with an average $40 \times 10^{-4}$.  Typical measurement uncertainties (excluding calibration errors) in a given pixel are $\sim 3$\% with the highest measurement uncertainties (and thus lowest signal to noise) of $\sim 20$\% found in the pixels on the outermost edges of the map. There are a few regions along the spiral arms in the north and southwest of the map where there is a small enhancement of the ratio, corresponding to peaks in the [C\,\textsc{ii}] emission.  In addition we see a lower ratio in the center of the galaxy, corresponding to a slight deficit of [C\,\textsc{ii}] emission.  The average [C\,\textsc{ii}]/$F_{\mathrm{TIR}}$ ratio along with the line/$F_{\mathrm{TIR}}$ ratio for each of the other far-infrared lines is shown in Table~\ref{tbl:line2tir}.  Not surprisingly, the next strongest ratio after the [C\,\textsc{ii}]/$F_{\mathrm{TIR}}$ ratio is the [O\,\textsc{ii}]63/$F_{\mathrm{TIR}}$ ratio.  The emission from these lines accounts for between 0.01\% and 0.4\% of the total infrared flux in this region of M51.

\begin{deluxetable}{lcc}
\tabletypesize{\small}
\tablecolumns{3}
\tablecaption{Line to total infrared flux ratio in M51\label{tbl:line2tir}}
\tablewidth{0pt}
\tablehead{
\colhead{Line} 	            & \colhead{$10^{-4}$ Line/$F_{\mathrm{TIR}}$\tablenotemark{a}} &
	\colhead{Area\tablenotemark{b} ($\sq\arcsec$)}}
  \startdata
 $[$C\,\textsc{ii}](158~$\mu$m) & $40 \pm 10$ & 21360 \\
 $[$N\,\textsc{ii}](122~$\mu$m) & $5 \pm 1$ & 17536 \\
 $[$O\,\textsc{i}](63~$\mu$m)   & $9 \pm 3$ & 17472 \\
 $[$O\,\textsc{i}](145~$\mu$m)  & $1.0 \pm 0.7$ & 3008 \\
 $[$O\,\textsc{iii}](88~$\mu$m) & $2 \pm 1$ & 2976 \\
 \enddata
 \tablenotetext{a}{Average spectral line flux divided by the total infrared flux, as calculated using our 5$\sigma$-cut maps for M51.  The uncertainties shown are the standard deviations.}
 \tablenotetext{b}{The area over which each average is calculated.  The variations reflect the size differences in our maps between different fine structure lines.}
\end{deluxetable}

The global [C\,\textsc{ii}]/$F_{\mathrm{TIR}}$ ratio has been measured by numerous surveys of both nearby galaxies and high redshift sources to investigate the [C\,\textsc{ii}] deficit.  Studies of the [C\,\textsc{ii}]/$L_{\mathrm{FIR}}$ ratio in ultra luminous infrared galaxies (ULIRGs) show a deficit when compared to normal galaxies, with values of less than $5 \times 10^{-4}$ \citep[e.g.][]{1998ApJ...504L..11L,2003ApJ...594..758L}.  This is the so called [C\,\textsc{ii}] deficit.  Multiplying our [C\,\textsc{ii}]/$F_{\mathrm{TIR}}$ ratio by a factor of 1.3 to convert $F_{\mathrm{TIR}}$ to $F_{\mathrm{FIR}}$ \citep{2008A&A...479..703G}, we find our [C\,\textsc{ii}]/$F_{\mathrm{FIR}}$ range of $13 \times 10^{-4}$ to $130 \times 10^{-4}$ with an average $52 \times 10^{-4}$ is in good agreement with previous studies of the [C\,\textsc{ii}]/$F_{\mathrm{FIR}}$ ratio.  \citet{2001ApJ...561..203N} previously investigated M51 and found this ratio varied between $60 \times 10^{-4}$ and $140 \times 10^{-4}$.  \citet{1985ApJ...289..803S} find a global value for the [C\,\textsc{ii}]/$F_{\mathrm{FIR}}$ ratio within the Milky Way of $30 \times 10^{-4}$ and on smaller scales, \citet{1993ApJ...404..219S} find the same value for the Orion molecular cloud.  \citet{1985ApJ...291..755C} conducted a study of six gas-rich galaxies, including M51 and find the ratio is about $50 \times 10^{-4}$ for this sample. \citet{2001ApJ...561..766M} find ratios of greater than $20 \times 10^{-4}$ in two-thirds of their sample of 60 normal star forming galaxies, and \citet{2011ApJ...728L...7G} find a range of values between $1 \times 10^{-4}$ and $100 \times 10^{-4}$ for a sample of 44 AGN and starburst type galaxies from the SHINING survey.  On smaller scales, \citet{2002AJ....124..751C} investigated NGC~6946 and NGC~1313 and find the ratio for these two galaxies is $80 \times 10^{-4}$, while \citet{2013A&A...549A.118C} find the [C\,\textsc{ii}]/$F_{\mathrm{FIR}}$ ratio varies between roughly $10 \times 10^{-4}$ and $100 \times 10^{-4}$ for various regions within the starburst M82.  \citet{2013A&A...553A.114K} recently discovered a radial trend in the [C\,\textsc{ii}]/$F_{\mathrm{FIR}}$ ratio in M33, increasing from $80 \times 10^{-4}$ in the inner galaxy to $300 \times 10^{-4}$ at a distance of roughly 4.5~kpc from the center, and thus increasing with decreasing far-infrared flux.

We also compare our results to the empirically derived relation from \citet{2012ApJ...745..171S} predicting the global [C\,\textsc{ii}] luminosity given a galaxy's total infrared luminosity.  Using our total-infrared luminosity of $4.71 \times 10^{10} L_{\odot}$ and Equation~33 from \citet{2012ApJ...745..171S}, we calculate a global [C\,\textsc{ii}] luminosity of $7.5 \times 10^{7} L_{\odot}$, thus a [C\,\textsc{ii}]/$L_{\mathrm{FIR}}$ ratio of $16 \times 10^{-4}$.  This value in agreement with the lower range of our observed values.  Thus, we find that even though our [C\,\textsc{ii}]/$F_{\mathrm{FIR}}$ ratio is smaller (by up to a factor of two) in the nucleus than in the rest of M51, it is not as extreme as the low values seen in ULIRGs.

\citet{2001A&A...375..566N} calculated the line/$F_{\mathrm{TIR}}$ ratio for the same far-infared lines we measured here with PACS.  Their sample of galaxies, including normal, starburst and AGN types, shows line/$F_{\mathrm{TIR}}$ ratios consistent with those measured in M51.  Furthermore, our results for M51 are also consistent with the results \citet{2001ApJ...561..766M} found for their sample of normal galaxies (note we have increased their far-infrared flux values by a factor of 1.3 to approximate the total infrared flux).  We compare our results to both papers in Table~\ref{tbl:comp_line2tir}.  Our resolved average values for M51 fall within the range of ratios typically found in a variety of galaxy types on unresolved scales.

\begin{deluxetable}{lcccc}
\tabletypesize{\small}
\tablecolumns{5}
\tablecaption{line/$F_{\mathrm{TIR}}$ ratios in M51 compared to previous global surveys\label{tbl:comp_line2tir}}
\tablewidth{0pt}
\tablehead{
\colhead{Line}	            & \multicolumn{4}{c}{$10^{-4}$ Line/$F_{\mathrm{TIR}}$} \\
\colhead{} & \multicolumn{2}{c}{This work\tablenotemark{a}} & \colhead{\citet{2001A&A...375..566N}} & \colhead{\citet{2001ApJ...561..766M}} \\
\colhead{} & \colhead{Average} & \colhead{Range} & \colhead{} & \colhead{}}
  \startdata
 $[$C\,\textsc{ii}](158~$\mu$m) & $40 \pm 10$   & 9--100 & 7--50  & 2--100 \\
 $[$N\,\textsc{ii}](122~$\mu$m) & $5 \pm 1$     & 2--10 & 1--7   & 0.8--8 \\
 $[$O\,\textsc{i}](63~$\mu$m)   & $9 \pm 3$     & 4--30 & 4--40  & 5--30 \\
 $[$O\,\textsc{i}](145~$\mu$m)  & $1.0 \pm 0.7$ & 0.3--5 & \nodata & \nodata \\
 $[$O\,\textsc{iii}](88~$\mu$m) & $2 \pm 1$     & 0.9--8 & 1--10  & 2--20 \\
 \enddata
 \tablenotetext{a}{Note that only 5$\sigma$ detections are included; global values for M51 are likely to be lower.}
\end{deluxetable}

\subsection{Heating and cooling}
In Figure~\ref{fig:ratio_plots} (middle) we show a map of ([C\,\textsc{ii}]+[O\,\textsc{i}]63)/$F_{\mathrm{TIR}}$, which is considered a proxy for the total heating efficiency, $\epsilon$ \citep{1985ApJ...291..722T}.  The heating efficiency represents the fraction of energy from the interstellar FUV radiation field that is converted to gas heating through the photoelectric effect, divided by the fraction of its energy deposited in dust grains.  This ratio represents the heating efficiency within the context of gas heating in PDR regimes (see for example \citet{1985ApJ...291..722T}).  A comparison with the [C\,\textsc{ii}]/$F_{\mathrm{TIR}}$ ratio shows that adding the [O\,\textsc{i}]63 emission enhances some of the structure in the spiral arms.  We look at this ratio in more detail in Section~\ref{pdr_model}.

The [C\,\textsc{ii}]/[O\,\textsc{i}]63 ratio is also shown in Figure~\ref{fig:ratio_plots} (bottom).  Cooling via the [C\,\textsc{ii}] line is more efficient in lower density, lower temperature regimes while [O\,\textsc{i}]63 cooling dominates at higher densities and warmer temperatures \citep{1985ApJ...291..722T}.  In this figure, the central region shows a ratio lower than the rest of the galaxy by a factor of $\sim\,2 - 4$, corresponding to the strong [O\,\textsc{i}]63 emission in the center; however, the ratio remains greater than 1.0 everywhere.  There is a visible increase in the ratio to upwards of 6.0 along the inner part of the eastern spiral arm, spatially coincident with the decreasing [C\,\textsc{ii}] emission at the outer edge of the nuclear region.  The low ratio in the center of the galaxy might indicate that the gas is warmer and/or more dense and cooling by the [O\,\textsc{i}]63 line becomes more important.  Typical measurement uncertainties are about 8\% with the noisiest pixels lying around the edge of the map having uncertainties of up to $\sim 24$\%.

The ratio of the two [O\,\textsc{i}] lines, [O\,\textsc{i}]145/[O\,\textsc{i}]63, can probe the temperature in the range around $\sim 300$~K for optically thin neutral gas because the excitation energies, $\Delta E/k$, are 228~K and 325~K above the ground state for the [O\,\textsc{i}]63 and [O\,\textsc{i}]145 lines, respectively \citep{1985ApJ...291..722T, 1999ApJ...527..795K, 2001ApJ...561..766M, 2006A&A...446..561L}.  However, the [O\,\textsc{i}]63 line can become optically thick at a lower column density, than the [O\,\textsc{i}]145 line, boosting the ratio of the two [O\,\textsc{i}] lines for gas temperatures less than $\sim 1000$~K \citep{1985ApJ...291..722T}.  Using a 5$\sigma$ cutoff for both the [O\,\textsc{i}]63 and [O\,\textsc{i}]145 lines and noting that the [O\,\textsc{i}]145 was only mapped along a radial strip (see Figure~\ref{fig:pacs_spec_maps}), our measurements of the ratio reside primarily in the central region of the galaxy, along with a few pixels in the inner part of the northwestern spiral arm.  In the center region the average ratio is 0.08 with typical measurement uncertainties of $\sim 9$\%.  Taking the inverse of this value to obtain a [O\,\textsc{i}]63/[O\,\textsc{i}]145 ratio of $12.5 \pm 5.3$ and comparing it with Figure~4 from \citet{2006A&A...446..561L}, we find that the [O\,\textsc{i}]63 line is either optically thick with $T \gtrsim 200$~K and $n \gtrsim 10^{3}$~cm$^{-3}$, or optically thin and hot with $T \sim 4000$~K and a density of approximately $10^{3}$~cm$^{-3}$.

\begin{figure}
\includegraphics[width=\columnwidth]{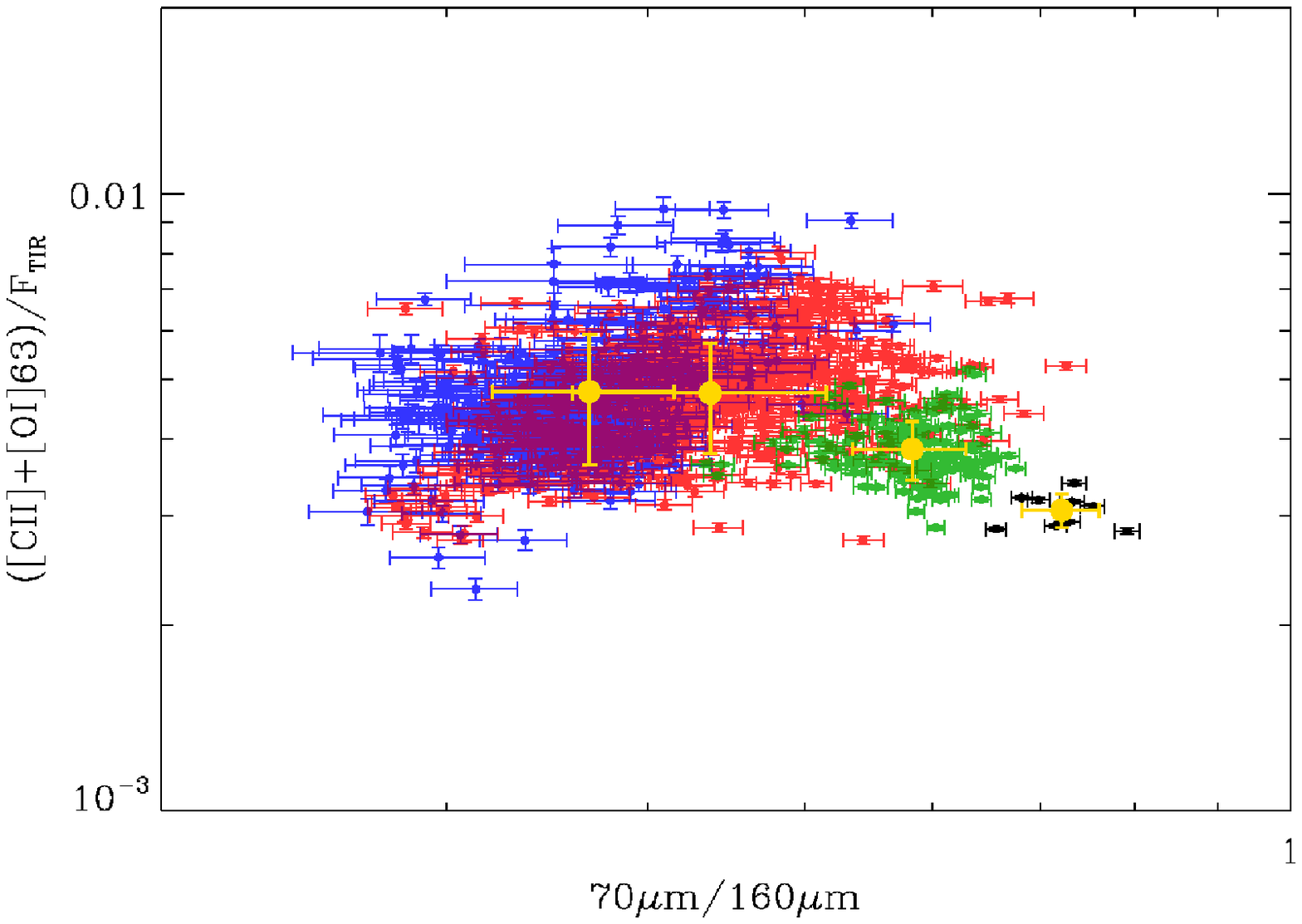}
\includegraphics[width=\columnwidth]{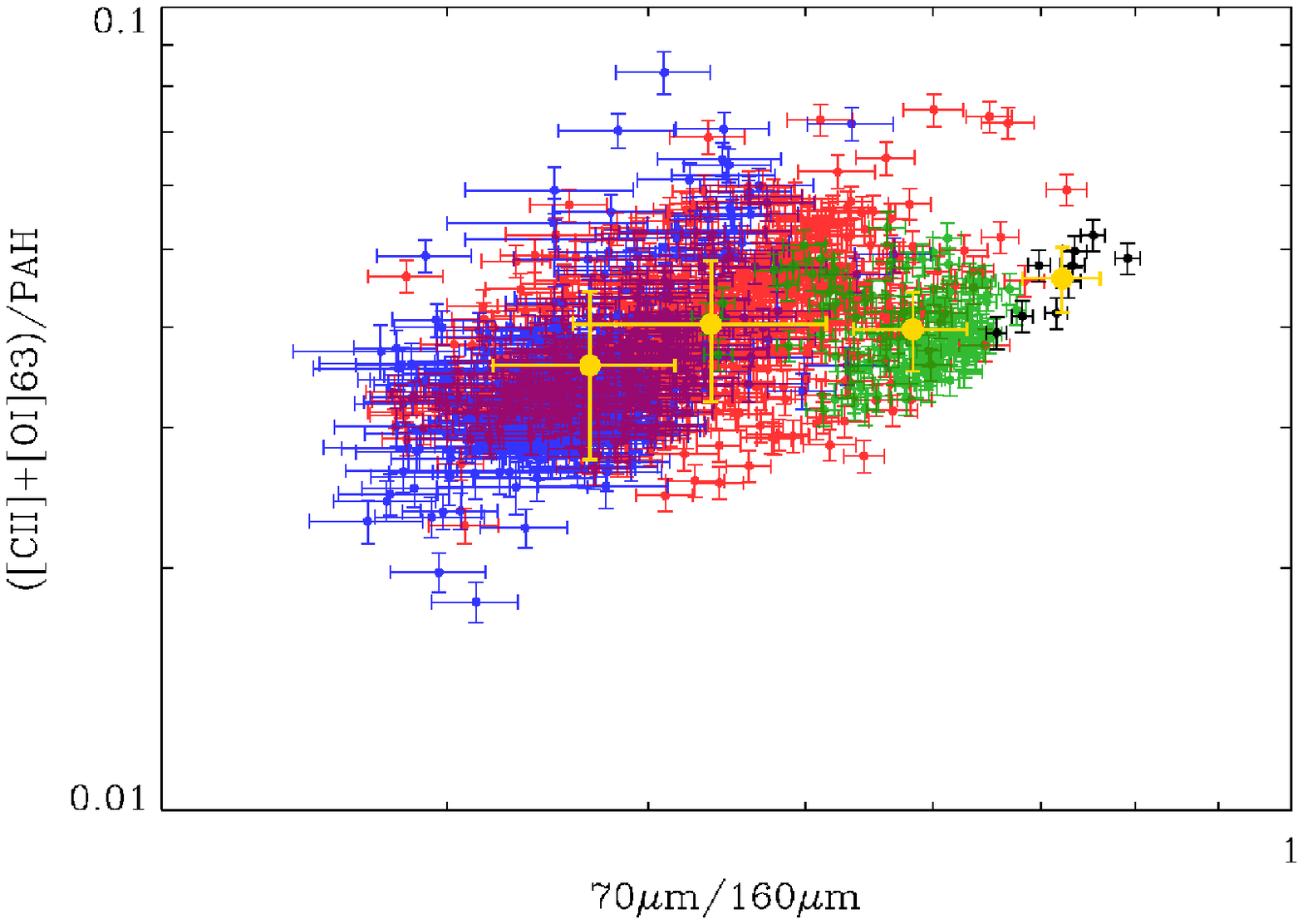}
\caption{\emph{top:} Total cooling ([C\,\textsc{ii}]+[O\,\textsc{i}]63) divided by the total infrared flux versus the PACS 70$\mu$m/160$\mu$m color.  \emph{bottom:} Total cooling ([C\,\textsc{ii}]+[O\,\textsc{i}]63) divided by the PAH emission, represented by the stellar subtracted IRAC 8~$\mu$m flux, versus the PACS 70$\mu$m/160$\mu$m color.  In both cases, one data point represents one pixel.  The pixels from the nucleus, center, arm and interarm regions are shown in black, green, red and blue, respectively.  The large yellow circles represent the mean for each region.}
\label{fig:heating_efficiency}
\end{figure}

We can also investigate the diagnostic plots ([C\,\textsc{ii}]+[O\,\textsc{i}]63)/$F_{\mathrm{TIR}}$ or ([C\,\textsc{ii}]+[O\,\textsc{i}]63)/$F_{\mathrm{PAH}}$ versus far-infrared color to look at the heating efficiency in more depth. Here we take the far-infared color 70$\mu$m/160$\mu$m, but typically the IRAS color 60$\mu$m/100$\mu$m is used.  Previous work has found a correlation between the heating efficiency and the infared color showing a decrease in heating efficiency with warmer colors, on both global and galactic scales \citep[e.g.][]{2001ApJ...561..766M,2012ApJ...747...81C}. This decrease has been attributed to warmer dust grains becoming increasingly positively charged when exposed to stronger radiation fields thus lowering the efficiency of the photoelectric effect.  It has also been shown that there is an even tighter correlation between heating efficiency and PAH emission, perhaps indicating that PAHs are the primary contributor to gas heating rather than dust grains in regions where [C\,\textsc{ii}] and [O\,\textsc{i}]63 are the primary coolants \citep{2012ApJ...747...81C, 2012A&A...548A..91L}.  In Figure~\ref{fig:heating_efficiency} we plot ([C\,\textsc{ii}]+[O\,\textsc{i}]63)/$F_{\mathrm{TIR}}$ as a function of 70$\mu$m/160$\mu$m (\emph{top}) and ([C\,\textsc{ii}]+[O\,\textsc{i}]63)/$F_{\mathrm{PAH}}$ as a function of color (\emph{bottom}).  In both cases, each data point corresponds to one pixel in our images and the colors correspond to the four different regions we break the galaxy into for our PDR analysis (see Section~\ref{pdr_model} and Figure~\ref{fig:regions} for more details).  We find that the heating efficiency as traced by ([C\,\textsc{ii}]+[O\,\textsc{i}]63)/$F_{\mathrm{TIR}}$ decreases by about a factor of 2 with increasing color as found by previous studies, which corresponds to a lower heating efficiency in the center of the galaxy than in the arm and interarm regions.  When we trace the heating efficiency by ([C\,\textsc{ii}]+[O\,\textsc{i}]63)/$F_{\mathrm{PAH}}$ we find a variation of approximately 30\% across the color space, in agreement with general trends found by \citet{2012A&A...548A..91L} in LMC~N11B and \citet{2012ApJ...747...81C} in NGC~1097 and NGC~4559.  However, this ratio does not vary significantly for the warmer dust, in contrast to the ([C\,\textsc{ii}]+[O\,\textsc{i}]63)/$F_{\mathrm{TIR}}$ ratio.

Our value for the total heating efficiency within PDRs as measured by ([C\,\textsc{ii}]+[O\,\textsc{i}]63)/$F_{\mathrm{TIR}}$ ranges between approximately $2 \times 10^{-3}$ and $5 \times 10^{-3}$.  The survey conducted by \citet{2001ApJ...561..766M} found a range of values between $10^{-3}$ and $10^{-2}$.  \citet{2012ApJ...747...81C} looked at NGC~1097 and NGC~4559 and find this proxy for the heating efficiency falls between approximately $2 \times 10^{-3}$ to $10^{-2}$; thus, our values for M51 fall within the lower range of their results.  In slightly lower metallicity environments the ratio is about the same at $2.77 \times 10^{-3}$ for NGC~4214 (metallicity of log(O/H)$+ 12 = 8.2$ \citep{2010A&A...518L..57C}) and $2 \times 10^{-3}$ in Haro~11 (metallicity 1/3 solar) \citep{2012A&A...548A..20C}.  On small scales, \citet{2012A&A...548A..91L} investigated the H\,\textsc{ii} region LMC-N11B (metallicity $\sim$1/2 solar), and found an average value of $\sim 5.5 \times 10^{-3}$ in PDR dominated regions, but the ratio decreased in regions dominated by ionized gas, a trend they attribute to contamination of the total infrared flux from the ionized gas where [C\,\textsc{ii}] and [O\,\textsc{i}]63 do not primarily emit.

Looking at the ([C\,\textsc{ii}]+[O\,\textsc{i}]63)/PAH ratio we find an average of $\sim0.01$, which is less than the value of 0.07 in LMC-N11B \citep{2012A&A...548A..91L}, as well as the range of 0.035--0.06 in NGC~1097 and NGC~4559 \citep{2012ApJ...747...81C}.  \citet{2012ApJ...751..144B} find that the ([C\,\textsc{ii}]+[O\,\textsc{i}]63)/PAH ratio varies from 0.03 to 0.1 and is approximately 50\% lower in the ring of NGC~1097 than in the nucleus.  However, \citet{2012ApJ...747...81C} note that for NGC~1097 and NGC~4559, using the IRAC 8$\mu$m map as a proxy for the total PAH emission overestimates (by about 10\%) the true total emission, estimated from running the \emph{Spitzer} IRS spectrum through PAHfit \citep{2007ApJ...656..770S}.  As a similar check for M51, we run the IRS spectrum for M51 produced by the SINGS team \citep{2003PASP..115..928K}, which is an average extracted from a region approximately $60\arcsec \times 35\arcsec$ centered on the nucleus, through PAHfit.  We then measure the total PAH intensity from within the same region using the IRAC 8~$\mu$m and compare it to the total calculated by PAHfit.  We find that for the region covered by IRS spectrum, the IRAC 8~$\mu$m map overestimates the total PAH intensity by a factor of $\sim3.5$.  Applying this correction to the ([C\,\textsc{ii}]+[O\,\textsc{i}]63)/PAH ratio in all four regions we investigate here will bring the data points in Figure~\ref{fig:heating_efficiency} (\emph{bottom}) up by a factor of 3.5, and thus in better agreement with the range of values determined by \citep{2012A&A...548A..91L}, \citet{2012ApJ...751..144B} and \citep{2012ApJ...747...81C}.  We note that our PAH map has not been corrected for the underlying dust continuum, which may partially explain the discrepancy between the PAH intensity measured by PAHfit and that measured by the IRAC 8~$\mu$m map.

\subsection{Ionized gas}\label{ionized_gas}
\subsubsection{Ionized gas contribution to [C\,\textsc{ii}] emission}\label{ionized_gas_contribution}
To compare our [C\,\textsc{ii}] map properly with the theoretical results of \citet{1999ApJ...527..795K, 2006ApJ...644..283K}, we need to correct for the fraction of [C\,\textsc{ii}] emission arising from ionized gas, as [C\,\textsc{ii}] emission can originate in both neutral and ionized gas.  We follow the method of \citet{2006ApJ...652L.125O} and use the diagnostic capabilities of the [N\,\textsc{ii}]122/[N\,\textsc{ii}]205 and [C\,\textsc{ii}]/[N\,\textsc{ii}]205 line ratios. [N\,\textsc{ii}] emission arises entirely from ionized gas because the ionization potential of N$^{+}$ is greater than 13.6~eV. In addition, the ratio of its two fine-structure lines is a sensitive probe of the gas density in H\,\textsc{ii}~regions. The critical densities of the [N\,\textsc{ii}]122 and [N\,\textsc{ii}]205 lines are 293 and 44~cm$^{-3}$, respectively, when $T_{e} = 8000$~K (commonly adopted for H\,\textsc{ii}~regions) \citep[][]{2006ApJ...652L.125O}.  Furthermore, at the same temperature, the [C\,\textsc{ii}] line has a critical density of 46~cm$^{-3}$ for collisions with electrons \citep[][]{2006ApJ...652L.125O} and so the [C\,\textsc{ii}]/[N\,\textsc{ii}]205 ratio is primarily dependent on the abundances of C$^{+}$ and N$^{+}$. A comparison of the theoretical ratio to the observed ratio at a specific electron density will determine the fraction of the [C\,\textsc{ii}] flux coming from ionized gas.  We compute the theoretical curves for the [N\,\textsc{ii}]122/[N\,\textsc{ii}]205 and [C\,\textsc{ii}]/[N\,\textsc{ii}]205 line ratios as a function of electron density using Solar gas phase abundances of C/H~$= 1.4 \times 10^{-4}$ and N/H~$= 7.9 \times 10^{-5}$ \citep{1996ARA&A..34..279S}, collision strengths for the [N\,\textsc{ii}] and [C\,\textsc{ii}] lines from \citet{2004MNRAS.348.1275H} and \citet{1992ApJS...80..425B}, respectively, and Einstein coefficients from \citet{1997A&AS..123..159G} and \citet{1998A&AS..131..499G} for the [N\,\textsc{ii}] and [C\,\textsc{ii}] transitions, respectively.  We choose to adopt Solar abundances here because \citet{2004AJ....128.2772G} showed that the C and N abundances in M51 are consistent with the Solar values within their uncertainties.  We note that while M51 has a slightly higher metallicity than solar \citep{2009ARA&A..47..481A} and has a slight decrease in metallicity with increasing radius \citep{2010ApJS..190..233M,2012ApJ...755..165M}, \citet{1999ApJ...513..168G} showed that the C/N abundance ratio is not affected by metallicity gradients in two nearby spirals, M101 and NGC~2403.  Thus, we believe the metallicity gradient will not strongly affect our results.

\begin{figure}
\includegraphics[width=\columnwidth]{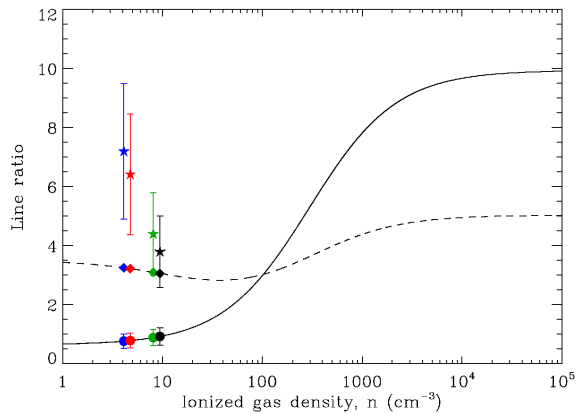}
\caption{A comparison of the average observed line ratio to the theoretical curves for each of the four regions we probe in M51.  The black solid line represents the theoretical curve for the [N\,\textsc{ii}]122/[N\,\textsc{ii}]205 line ratio while the black dashed line represents the theoretical curve for the [C\,\textsc{ii}]/[N\,\textsc{ii}]205 line ratio.  The pixels from the nucleus, center, arm and interarm regions are shown in black, green, red and blue, respectively.  The solid dots show where the observed [N\,\textsc{ii}]122/[N\,\textsc{ii}]205 line ratios for each region fall on the theoretical curve thus allowing us to determine the ionized gas density.  The diamonds (stars) show the theoretical (observed) values of the [C\,\textsc{ii}]/[N\,\textsc{ii}]205 line ratio at the inferred ionized gas density.  The error bars on the observed line ratios include calibration uncertainties.}
\label{fig:line_ratio_compare}
\end{figure}

To measure the observed [N\,\textsc{ii}]122/[N\,\textsc{ii}]205 and [C\,\textsc{ii}]/[N\,\textsc{ii}]205 line ratios, we convolved our [C\,\textsc{ii}] and [N\,\textsc{ii}]122 maps to the resolution of the [N\,\textsc{ii}]205 map ($\sim$17$\arcsec$) and then calculated the ratios.  Our [N\,\textsc{ii}]205 map only contains a small number of finite pixels; thus the resulting ratio maps give us an estimate of [N\,\textsc{ii}]122/[N\,\textsc{ii}]205 and [C\,\textsc{ii}]/[N\,\textsc{ii}]205 (at positions observed in [N\,\textsc{ii}]205) that we can use to estimate the electron density and fraction of [C\,\textsc{ii}] emission from ionized gas in each region.  Comparing our observed ratios in each region to the theoretical curves (Figure~\ref{fig:line_ratio_compare}) we find that the fraction of the [C\,\textsc{ii}] emission in our observations coming from ionized gas is $0.8 \pm 0.2$, $0.7^{+0.3}_{-0.2}$, $0.5^{+0.3}_{-0.2}$ and $0.5^{+0.2}_{-0.1}$ for the nucleus, center, arm and interarm regions, respectively (Table~\ref{table:ratio_properties}).  Calibration uncertainties are included and have the effect of shifting all our ratios up or down, but the trend from region to region will remain the same.

\begin{deluxetable}{lccccc}
\tabletypesize{\small}
\tablecolumns{6}
\tablecaption{[N\,\textsc{ii}] line ratios and ionized fraction of [C\,\textsc{ii}]\label{table:ratio_properties}}
\tablewidth{0pt}
\tablehead{
\colhead{Region} & \colhead{Observed} & \colhead{$n_{e}$\tablenotemark{a} (cm$^{-3}$)}
	& \colhead{Predicted} & \colhead{Observed} & \colhead{Ionized fraction} \\
    & \colhead{$[$N\,\textsc{ii}]$_{122}$/[N\,\textsc{ii}]$_{205}$} &
    & \colhead{$[$C\,\textsc{ii}]$^{ionized}_{158}$/[N\,\textsc{ii}]$_{205}$\tablenotemark{b}}
    & \colhead{$[$C\,\textsc{ii}]$_{158}$/[N\,\textsc{ii}]$_{205}$} & \colhead{of [C\,\textsc{ii}]}}
 \startdata
 nucleus  & $0.9^{+0.3}_{-0.2}$ & $9^{+10}_{-7}$   & $3.1^{+0.3}_{-0.2}$ & $4 \pm 1$ & $0.8^{+0.2}_{-0.2}$ \\
 center   & $0.9^{+0.3}_{-0.2}$ & $8^{+10}_{-6}$   & $3.1^{+0.3}_{-0.2}$ & $4 \pm 1$ & $0.7^{+0.3}_{-0.2}$ \\
 arm      & $0.78^{+0.25}_{-0.08}$ & $5^{+8}_{-3}$ & $3.2^{+0.1}_{-0.2}$ & $6 \pm 2$ & $0.5^{+0.3}_{-0.2}$ \\
 interarm & $0.76^{+0.24}_{-0.06}$ & $4^{+8}_{-2}$ & $3.2^{+0.1}_{-0.3}$ & $7 \pm 2$ & $0.5^{+0.2}_{-0.1}$ \\
 \enddata
 \tablecomments{These are average values for each of the four regions, and calibration uncertainties are included.}
 \tablenotetext{a}{Derived from [N\,\textsc{ii}]122/[N\,\textsc{ii}]205; see text.}
 \tablenotetext{b}{Derived from $n_{e}$ and theoretical prediction of [C\,\textsc{ii}]/[N\,\textsc{ii}]205 ratio in ionized gas; see text.}
\end{deluxetable}

\subsubsection{Ionized Gas Characteristics}
The fraction of N$^{+}$ originating from diffuse ionized gas versus that originating from H\textsc{ii}~regions has been discussed at length in the literature.  A set of models for H\,\textsc{ii} regions by \citet{1985ApJS...57..349R} as well as a later model by \citet{1994ApJ...420..772R} show that the ratio between the two [N\,\textsc{ii}] lines, [N\,\textsc{ii}]122/[N\,\textsc{ii}]205, varies from 3 for $n_{\mathrm{e}} \sim 100~\mathrm{cm}^{-3}$ to 10 for $n_{e} \gtrsim 10^{3}~\mathrm{cm}^{-3}$.  When $n_{e} \ll n_{\mathrm{critical}}$, the theoretical value for this ratio is 0.7 \citep{1991ApJ...381..200W,1994ApJ...434..587B}.  Observations of [N\,\textsc{ii}]122 and [N\,\textsc{ii}]205 in the Milky Way show that the [N\,\textsc{ii}]122/[N\,\textsc{ii}]205 falls between 1.0 and 1.6 \citep{1991ApJ...381..200W}.  Comparison between the observed and theoretical ratios implies that between 60 and 87\% of the [N\,\textsc{ii}] emission in the Milky Way as measured with the Cosmic Background Explorer (COBE) comes from diffuse gas while the remainder comes from H\,\textsc{ii} regions with $n_{\mathrm{e}} \sim 100~\mathrm{cm}^{-3}$.  \citet{1994ApJ...434..587B} re-examine the data presented in \citet{1991ApJ...381..200W} and find [N\,\textsc{ii}]122/[N\,\textsc{ii}]205 = $0.9 \pm 0.1$ assuming a constant ratio, but also note that their data are better fit when the ratio is allowed to vary.  Our observations show that [N\,\textsc{ii}]122/[N\,\textsc{ii}]205 ranges between $0.76^{+0.24}_{-0.06}$ and $0.92^{+0.3}_{-0.2}$, thus approaching the theoretical lower limit of 0.7, implying that much of the ionized gas is diffuse.  Nonetheless, our mean values of the [N\,\textsc{ii}] line ratios are consistent with those of the Milky Way.

Our values for the fraction of [C\,\textsc{ii}] emission coming from PDRs are lower than those determined in other nearby galaxies, as well as previous calculations for M51.  Furthermore, a gradient in this fraction has not previously been observed in other galaxies, and is an important result.  Given that we believe the Seyfert nucleus is not the major source of gas excitation in the center (see Section~\ref{seyfert_influence}), the high fraction of ionized gas likely indicates that there are more massive stars per unit volume providing the ionizing photons.

The [N\,\textsc{ii}]205 line is difficult to detect, and so it is often necessary to rely on the [C\,\textsc{ii}]/[N\,\textsc{ii}]122 line ratio to determine the ionized gas fraction in the [C\,\textsc{ii}] emission.  This ratio requires knowledge about the density of the gas, which can sometimes be difficult to obtain.  However, in the absence of [N\,\textsc{ii}]205 observations it is often the best way to estimate the ionized gas contribution to the observed [C\,\textsc{ii}] emission.  \citet{2001ApJ...561..766M} used the [C\,\textsc{ii}]/[N\,\textsc{ii}]122 emission and the Galactic value of the [N\,\textsc{ii}]122/[N\,\textsc{ii}]205 ratio to estimate that about 50\% of the observed [C\,\textsc{ii}] emission in their sample of galaxies orginiated in PDRs, in agreement with our results for the arm and interarm regions.  However, \citet{2005A&A...441..961K} used a similar method to determine that 70 to 85\% of [C\,\textsc{ii}] emission comes from PDRs in M51 and M83.  This is a much higher value than what we determine using the [N\,\textsc{ii}]205 line for M51, especially in the center and nuclear regions where we find a large fraction of [C\,\textsc{ii}] emission is coming from ionized gas.  In both cases the correction was applied globally, but our results demonstrate that this method may not accurately correct the observed [C\,\textsc{ii}] emission and thus lead to incorrect results when comparing observations to PDR models.

The [O\,\textsc{iii}] can also be used to probe the ionized gas.  One diagnostic used for high redshift galaxies is the ratio of [O\,\textsc{iii}]/[N\,\textsc{ii}]122.  \citet{2011ApJ...740L..29F} showed that this ratio can constrain the hardness of the UV radiation field as the ionization potentials of N and O$^{+}$ are 14.5 and 35~eV, respectively, while the [N\,\textsc{ii}]122 and [O\,\textsc{iii}] lines have critical densities of 310 and 510~cm$^{-3}$, respectively.  This means the line ratio is relatively constant as a function of gas density.  If the emission arises from H\,\textsc{ii} regions, then the ratio gives an indication of the effective stellar temperature of the ionizing source(s).  If the emission comes from an AGN, particularly from the narrow line region (NLR), the ratio indicates the value of the ionization parameter, $U$.  This parameter represents the photon density of the incident radiation field on a molecular cloud divided by the gas density of the cloud, and gives an indication of the amount of dust absorbing the ionizing flux \citep{2003ApJ...594..758L,2009ApJ...701.1147A}.

In their Figure~1, \citet{2011ApJ...740L..29F} compare their observed [O\,\textsc{iii}]/[N\,\textsc{ii}]122 line ratio to that predicted for H\,\textsc{ii} regions using the model of \citet{1985ApJS...57..349R}, as well as the ratio predicted for an NLR within an AGN using the model of \citet{2004ApJS..153....9G}. For their high redshift galaxy, the ratio satisfies either model as well as a combination of the two.  We find that for the nucleus region of M51, which represents the area covered by the PACS point spread function, the [O\,\textsc{iii}]/[N\,\textsc{ii}]122 ratio is $0.33 \pm 0.05$, while for the somewhat larger `center' region we measure $0.23 \pm 0.05$.  Following the method of \citet{2011ApJ...740L..29F}, we compare our observed ratios to model predictions.  Assuming the emission is from an H\,\textsc{ii} region, our results suggest the most luminous stars are B0 \citep{1996ApJ...460..914V} for both the nucleus and center regions.  On the other hand if we assume the emission is from the AGN, our results imply a small ionization parameter of $10^{-4}$ to $10^{-3.5}$, which means the incident ionizing flux is weak.  A comparison of our average [O\,\textsc{iii}]/$F_{\mathrm{TIR}}$ ratio to model predictions of the ratio for AGN and starburst galaxies from \citet{2009ApJ...701.1147A} shows our ratio of $2\pm1$ is also consistent with a low value of $U$.  Furthermore, \citet{2004A&A...414..825S} conducted a survey of galaxies with low luminosity nuclei using IRS LWS spectroscopy, including M51. They determined that the flux of the [O\,\textsc{iv}] line, which is only excited in AGN due to the strong ionization potential of O$^{++}$ (55~eV), is lower in the nucleus of M51 than in typical AGN.  This result also indicates weak activity in the nucleus of M51. Thus, we conclude that the Seyfert nucleus in M51 is not significantly affecting the excitation of the gas.


\section{PDR modelling of observations}\label{pdr_model}
\begin{figure}[!h]
\includegraphics[width=7.0cm]{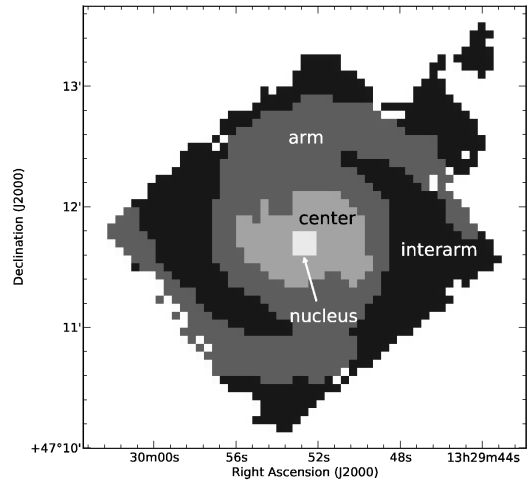}
\caption{A schematic of the four regions into which we divide M51 for our analysis.}
\label{fig:regions}
\end{figure}

We compare our observed line ratios to the PDR model of \citet{1999ApJ...527..795K,2006ApJ...644..283K}.  This model probes PDR regions with two free parameters, namely the density of hydrogen nuclei, $n$, and the strength of the FUV radiation field incident on the PDR, $G_{0}$.  \citet{1999ApJ...527..795K} consider a density range of $10^{1}\,\mathrm{cm}^{-3} \le n \le 10^{7}\,\mathrm{cm}^{-3}$ and a FUV radiation field range of $10^{-0.5} \le G_{0} \le 10^{6.5}$.  Here we look at the inner part of M51 by carrying out a pixel-by-pixel comparison between the model and our observations, and consider pixels within four regions, namely the ``nucleus'', the ``center'', the ``arm'' and the ``interarm'' regions.  These regions were distinguished using flux cutoffs in our total infrared flux map (Figure~\ref{fig:Ltir}) to isolate the nucleus from the rest of the center region and the spiral arms from the interarm regions.  Breaking the galaxy down into four distinct subregions allows us to probe the gas in different environments within the galaxy.  These regions are outlined in Figure~\ref{fig:regions} with cutoff maximum fluxes of $9.6 \times 10^{-7}$ and $3.7 \times 10^{-5}$~W~m$^{-2}$~sr$^{-1}$ for the interarm and arm regions, respectively.  The center region consists of everything above $3.7 \times 10^{-5}$~W~m$^{-2}$~sr$^{-1}$ except for the central nine pixels which comprise the nucleus.  We note that with a pixel scale of 4$\arcsec$ in our maps, each pixel is not independent from its neighbours.  In Table~\ref{table:flux_values} we list the average integrated intensity measured in each region for each of the five far-infrared lines that we have observed. 

\begin{deluxetable}{lccccc}
\tabletypesize{\small}
\tablecolumns{6}
\tablecaption{Average intensity of fine-structure lines by region\label{table:flux_values}}
\tablewidth{0pt}
\tablehead{
\colhead{Region} & \multicolumn{5}{c}{Integrated Intensity (W~m$^{-2}$~sr$^{-1}$)} \\
          & \colhead{[C\,\textsc{ii}](158~$\mu$m)} & \colhead{[N\,\textsc{ii}](122~$\mu$m)}
          & \colhead{[O\,\textsc{i}](63~$\mu$m)} & \colhead{[O\,\textsc{i}](145~$\mu$m)} & \colhead{[O\,\textsc{iii}](88~$\mu$m)}}
 \startdata
 nucleus\tablenotemark{a}  & $(1.4 \pm 0.1) \times 10^{-8}$  & $(3.2 \pm 0.2) \times 10^{-9}$
 	& $(8 \pm 1) \times 10^{-9}$      & $(7 \pm 2) \times 10^{-10}$ & $(1.0 \pm 0.1) \times 10^{-9}$ \\
 center\tablenotemark{b}   & $(1.1 \pm 0.2) \times 10^{-9}$  & $(1.7 \pm 0.4) \times 10^{-10}$
 	& $(2.9 \pm 0.7) \times 10^{-10}$ & $(3 \pm 1) \times 10^{-11}$ & $(8 \pm 2) \times 10^{-11}$ \\
 arms\tablenotemark{c}     & $(1.2 \pm 0.5) \times 10^{-10}$ & $(1.4 \pm 0.9) \times 10^{-11}$
 	& $(3 \pm 1) \times 10^{-11}$     & $(4 \pm 1) \times 10^{-11}$ & $(8 \pm 2) \times 10^{-11}$ \\
 interarm\tablenotemark{d} & $(5 \pm 2) \times 10^{-11}$     & $(8 \pm 3) \times 10^{-12}$
 	& $(1.8 \pm 0.5) \times 10^{-11}$ & $(3.2 \pm 0.8) \times 10^{-11}$ & $(1.7 \pm 0.4) \times 10^{-10}$ \\
 \enddata
 \tablecomments{Average integrated intensity measured in the four different regions for each of the far-infrared fine structure lines, using our maps with a 5$\sigma$ cut applied.  The uncertainties shown are the standard deviations.}
 \tablenotetext{a}{The number of pixels included in this measurement is 9 for all five lines.}
 \tablenotetext{b}{The number of pixels included in this measurement is 136 for the [C\,\textsc{ii}], [N\,\textsc{ii}]122 and [O\,\textsc{i}]63 lines, 95 for the [O\,\textsc{i}]145 line and 93 for the [O\,\textsc{iii}] line.}
 \tablenotetext{c}{The number of pixels included in this measurement is 583, 560, 553, 46, and 61 for the [C\,\textsc{ii}], [N\,\textsc{ii}]122, [O\,\textsc{i}]63, [O\,\textsc{i}]145 and [O\,\textsc{iii}] lines, respectively.}
 \tablenotetext{d}{The number of pixels included in this measurement is 607, 391, 394, 38, and 23 for the [C\,\textsc{ii}], [N\,\textsc{ii}]122, [O\,\textsc{i}]63, [O\,\textsc{i}]145, and [O\,\textsc{iii}] lines, respectively.}
\end{deluxetable}

\begin{figure*}
 \includegraphics[width=\columnwidth]{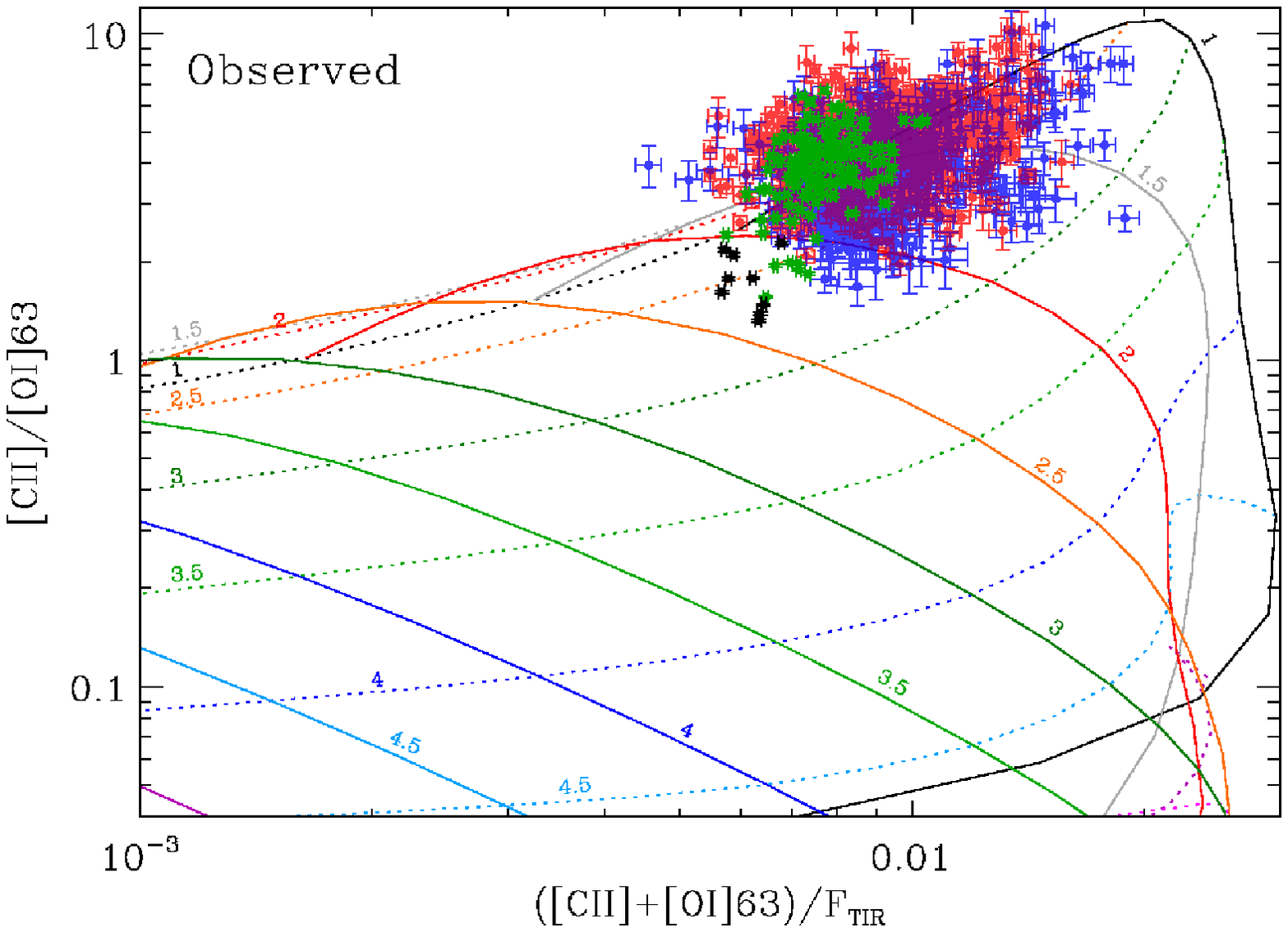}
 \includegraphics[width=\columnwidth]{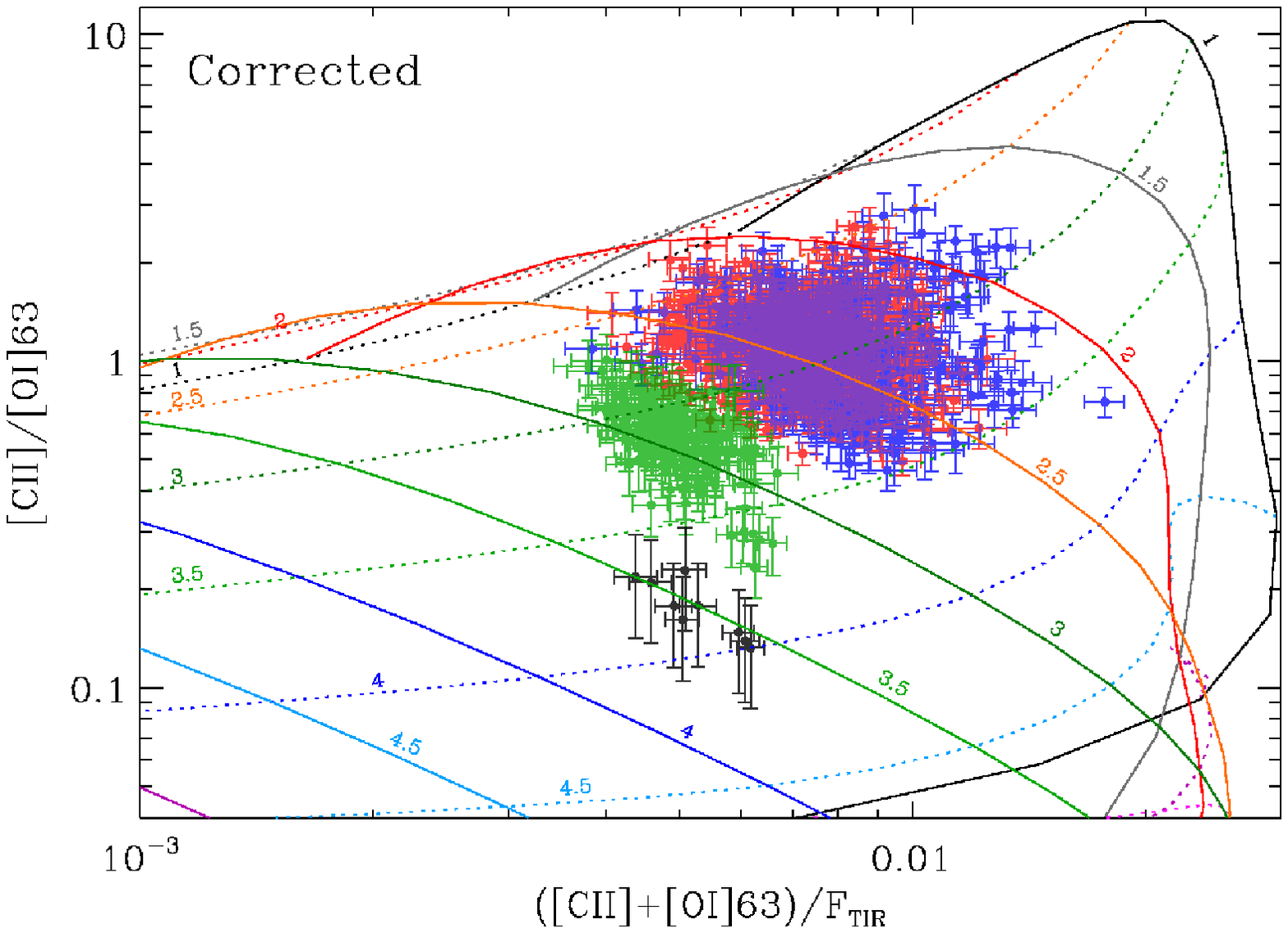}
\caption{Our observed data are overlaid on the PDR model grid of lines of constant log($n/\mathrm{cm}^{-3}$) (dotted lines) and log$G_0$ (solid lines).  One data point represents one pixel.  The pixels from the nucleus, center, arm and interarm regions are shown in black, green, red and blue, respectively.  \emph{left:}[C\,\textsc{ii}]/[OI]63 vs. ([C\,\textsc{ii}]+[OI]63)/$F_{\mathrm{TIR}}$ prior to removing the fraction of [C\,\textsc{ii}] emission from ionized gas. \emph{right}: [C\,\textsc{ii}]/[O\,\textsc{i}]63 vs. ([C\,\textsc{ii}]+[O\,\textsc{i}]63)/F$_{TIR}$ with the [C\,\textsc{ii}] emission corrected to removed the fraction originating in ionized gas, and the [O\,\textsc{i}]63 emission corrected for an ensemble of clouds (see text).}
\label{fig:CIIonOIvsCIIOIonLfir}
\end{figure*}

In Figure~\ref{fig:CIIonOIvsCIIOIonLfir} we show the [C\,\textsc{ii}]/[O\,\textsc{i}]63 ratio versus the ([C\,\textsc{ii}]+[O\,\textsc{i}]63)/$F_{\mathrm{TIR}}$ ratio for M51 overlaid on the parameter space defined by lines of constant log($n/\mathrm{cm}^{-3}$) (dotted lines) and log$G_{0}$ (solid lines) adapted from plots in \citet{1999ApJ...527..795K}.  The authors of that paper note that for extragalactic sources it is recommended that the observed total infrared flux be reduced by a factor of two to account for the (optically thin) infrared continuum flux coming from both the front and back sides of the cloud, whereas the model assumes emission is only coming from the front side of the cloud, just as the fine structure lines do.  Here we have applied this correction to our observed total infrared flux in order to compare it properly with the PDR model.  However, in this plot we have not yet corrected for the fraction of [C\,\textsc{ii}] emission arising from ionized gas (see Section~\ref{ionized_gas} for details).  We also note that the $F_{\mathrm{TIR}}$ we use for our comparison to the \citet{1999ApJ...527..795K} model is equivalent to their bolometric far-infrared flux.

With the exception of the nucleus, all of the data points tend to cluster around one locus and approximately one third of the pixels fall outside of the parameter space covered by the models.  The [C\,\textsc{ii}]/[O\,\textsc{i}]63 vs. ([C\,\textsc{ii}]+[O\,\textsc{i}]63)/$F_{\mathrm{TIR}}$ parameter space actually provides two possible model solutions.  One is a low-$G_0$, high-$n$ regime and the other is a regime with more moderate values for both parameters.  We do not display the low-$G_0$, high-$n$ solutions as we can eliminate these solutions by considering the number of clouds emitting within our beam.  When we compare the model predicted [C\,\textsc{ii}] emission based on the values of $G_0$ and $n$ to our observed [C\,\textsc{ii}] emission, we find that we would require a filling factor (i.e. the number of PDR regions) of upwards of $10^{3}$, which is an unreasonably large number of clouds along the line of sight.  This is the same reasoning used by \citet{2005A&A...441..961K} to eliminate the low-$G_0$, high-$n$ solution.  Thus, for the remaining discussion we consider only the moderate $n$ and $G_{0}$ solutions.

\subsection{Adjustments to the [C\,{\footnotesize II}] and [O\,{\footnotesize I}]63 lines}\label{pdr_model_adjustments}
A proper comparison to the PDR model of \citet{1999ApJ...527..795K} requires us to make two adjustments.  The first is to remove the fraction of [C\,\textsc{ii}] emission from ionized gas, as the model only applies to [C\,\textsc{ii}] emission from PDRs.  We use the results from Section~\ref{ionized_gas_contribution} to correct our [C\,\textsc{ii}] map.

The second correction is applied to the [O\,\textsc{i}]63 map.  The \citet{1999ApJ...527..795K} model is a plane-parallel slab that only experiences an incident radiation field on one side, which is the same side from which we observe emission from the far-infrared cooling lines.  In the case of M51, there are many clouds within a given PACS beam and the irradiated side of an individual cloud is not always oriented such that it is facing us.  \citet{1999ApJ...527..795K} caution that the velocity dispersion for such an ensemble of clouds combined with the assumption that the [O\,\textsc{i}]63 line will become optically thick much faster than either the [C\,\textsc{ii}] line or the total infrared flux means we observe only [O\,\textsc{i}]63 flux emitted from clouds with their front (lit) sides facing towards us, but we will observe [C\,\textsc{ii}] and total infrared flux from all clouds.  Thus, we only see about half of the total [O\,\textsc{i}]63 emission from all PDRs within our beam, and as a result we multiply the observed [O\,\textsc{i}]63 emission by a factor of two for comparison with the PDR model.

The resulting parameter space after the [C\,\textsc{ii}] and [O\,\textsc{i}]63 results have been corrected is shown in Figure~\ref{fig:CIIonOIvsCIIOIonLfir} (\emph{right}), and the values for $n$ and $G_0$ are presented in Table~\ref{table:pdr_model_results}.  Comparing the two panels of Figure~\ref{fig:CIIonOIvsCIIOIonLfir}, we can see that after applying the appropriate corrections to the [C\,\textsc{ii}] and [O\,\textsc{i}]63 emission there is a general shift of pixels to lower values of ([C\,\textsc{ii}]+[O\,\textsc{i}]63)/$F_{\mathrm{TIR}}$.  In addition, pixels from the center and nuclear regions have decreased more in [C\,\textsc{ii}]/[O\,\textsc{i}]63, as expected.  Looking at Table~\ref{table:pdr_model_results}, we see an increase in $n$ by approximately an order of magnitude for all four regions after we apply our corrections. Likewise, log$G_0$ increases by $\sim$ 1.5.

\begin{deluxetable}{lccccc}
\tabletypesize{\small}
\tablecolumns{5}
\tablecaption{Properties of the gas derived from the PDR model\label{table:pdr_model_results}}
\tablewidth{0pt}
\tablehead{
\colhead{Case} & \colhead{Region} & \colhead{log($n$/cm$^{-3}$)} & \colhead{log$G_{0}$}
	& \colhead{$T$ (K)}}
 \startdata
 uncorrected\tablenotemark{a} & nucleus  & 2.25-2.75  & 2.0-2.5    & 170-320 \\
             				  & center   & 1.5-2.75   & 1.0-2.5    & 70-1070 \\
             				  & arms     & 1.5-3.0    & 1.0-2.25   & 60-820  \\
             				  & interarm & 1.5-3.25   & 0.75-2.25  & 50-820  \\
 \hline
 corrected\tablenotemark{b}   & nucleus  & 3.5-4.25   & 3.25-4.0   & 240-475 \\
             				  & center   & 2.5-4.0    & 2.5-3.5    & 170-680 \\
             				  & arms     & 2.0-3.75   & 1.75-3.0   & 100-760 \\
             				  & interarm & 2.25-3.75  & 1.5-3.0    & 80-550  \\
 \enddata
 \tablenotetext{a}{The uncorrected case includes all of the observed [C\,\textsc{ii}] emission.}
 \tablenotetext{b}{The corrected case includes only [C\,\textsc{ii}] emission from neutral gas, and the [O\,\textsc{i}]63 has been increased by a factor of two as described in Section~\ref{pdr_model_adjustments}.}
\end{deluxetable}

We use our derived values for $n$ and $G_{0}$ to determine the range in the surface temperature of the gas using Figure~1 from \citet{1999ApJ...527..795K}.  The temperatures for the uncorrected [C\,\textsc{ii}] and [O\,\textsc{i}]63 observations as well as for the full corrected case are given in Table~\ref{table:pdr_model_results}.  In general, the cloud surface temperatures increase after the corrections are applied, while there is a decrease in temperature from the central region to the interarm region.

\subsection{Possible contamination from the Seyfert Nucleus}\label{seyfert_influence}
The [O\,\textsc{i}]63 emission peaks in the nucleus of M51, and as a result [C\,\textsc{ii}]/[O\,\textsc{i}]63 is lower in the center than in the rest of the galaxy.  In addition, the [O\,\textsc{iii}] line is also brightest in the nucleus and center regions, thus indicating higher densities and warmer temperatures in the center of the galaxy.  This is consistent with the results of our PDR modelling, but it is also possible that some of the [O\,\textsc{i}]63 emission from the nucleus is not arising from starlight but rather from shock heating.  Once the temperature of the gas behind a shock front has cooled down to below 5000~K, the [O\,\textsc{i}]63 line dominates the cooling for densities below $10^{5}$~cm$^{-3}$ \citep{1989ApJ...342..306H}.  If indeed some of the [O\,\textsc{i}]63 emission did arise from shocks rather than stellar radiation in PDRs, then this correction would mean that the [C\,\textsc{ii}]/[O\,\textsc{i}]63 ratio attributable to PDRs would increase in the center of the galaxy. Thus, the data point(s) in Figure~\ref{fig:CIIonOIvsCIIOIonLfir} would shift upwards and to the left in the parameter space, thus decreasing the gas density, and possibly the value of $G_{0}$ as well, although quantifying this effect would be difficult.  As a result the gradient we observe in density and radiation field due to PDRs would become less prominent.  We also note here that some of the total infrared flux may originate in non-PDR regions such as H\textsc{ii} regions.  Reducing this flux would move our data points higher in Figure~\ref{fig:CIIonOIvsCIIOIonLfir}.

\subsection{Implications of PDR model results}
The range of densities, temperatures and FUV radiation field strengths presented in Table~\ref{table:pdr_model_results} agree with the range of temperatures that we obtain for the [O\,\textsc{i}]63/[O\,\textsc{i}]145 line ratio if the [O\,\textsc{i}]63 line is optically thick ($T \gtrsim 200$~K).  In calculating these results, we have assumed that the corrections we applied to our observations of the fine structure lines (i.e. removing the [C\,\textsc{ii}] emission originating from diffuse ionized gas and accounting for [O\,\textsc{i}]63 emission escaping away from our line of sight) and the total infrared flux (reducing the total observed flux by a factor of two to remove emission originating from the back side of or even beyond the cloud) are correct for comparing our observations to the PDR model.  However, it is possible that some of the remaining $F_{\mathrm{TIR}}$ emission originates from regions other than PDRs (such as H\,\textsc{ii} regions or the diffuse ISM), meaning less than half of the total observed flux comes from PDRs.  In such a case, the data points in the ``Corrected'' panel of Figure~\ref{fig:CIIonOIvsCIIOIonLfir} would shift to the right, thus resulting in lower values of $G_{0}$ and higher densities.  Looking at this figure we see that we can only reduce the fraction of $F_{\mathrm{TIR}}$ originating in PDRs to approximately 1/8 of the total observed infrared flux before our data would no longer be consistent with the PDR model.  Thus, our data and analysis are consistent with at least 12\% and at most 50\% of $F_{\mathrm{TIR}}$ arising from PDRs in M51.

Our values of $n$ and $G_{0}$ are in good agreement with previous extragalactic surveys. Furthermore, our results agree with resolved studies of individual galaxies.  However, none of these studies shows a decreasing trend with radius such as we show here, not only because of resolution constraints, but also because most previous work has not looked at variations with region within an individual galaxy.  \citet{2005A&A...441..961K} searched within the center of M51 as well as two positions in the spiral arms with ISO; however, they were unable to resolve any differences between the three locations.

We see that the arm and interarm regions have approximately the same range of $G_{0}$ and $n$ despite lower star formation rate surface densities in the interarm region.  This may indicate that molecular clouds and star formation are similar in the arm and interarm regions, but there are just fewer clouds per unit area in the interarm regions.  To confirm this result we have calculated the mean values of [C\,\textsc{ii}]/[O\,\textsc{i}]63 and ([C\,\textsc{ii}]+[O\,\textsc{i}]63)/$F_{\mathrm{TIR}}$ for each region and compared them to the PDR model.  The results are summarized in Table~\ref{table:pdr_model_result_mean}.  We also compare these results with surveys and individual galaxies in Table~\ref{pdr_comparison}.

\begin{deluxetable}{lcccc}
\tabletypesize{\small}
\tablecolumns{5}
\tablecaption{Properties of the gas from the PDR model and mean line ratios\label{table:pdr_model_result_mean}}
\tablewidth{0pt}
\tablehead{
\colhead{Region}  & \colhead{Average\tablenotemark{a}} & \colhead{Average\tablenotemark{a}} & \colhead{log($n$/cm$^{-3}$)}
		& \colhead{log$G_{0}$} \\
                  & \colhead{($[\mathrm{CII}]+[\mathrm{OI}]$(63~$\mu$m))/$F_{\mathrm{TIR}}$}
  				  & \colhead{$[\mathrm{CII}]/[\mathrm{OI}]$(63~$\mu$m)} &  & }
 \startdata
 nucleus  & $(5.3 \pm 0.2) \times 10^{-3}$   & $0.18 \pm 0.01$ & 3.75-4.0           & 3.25-3.75 \\
 center   & $(5.01 \pm 0.05) \times 10^{-3}$ & $0.60 \pm 0.01$ & 3.0-3.25           & 2.75-3.0  \\
 arms     & $(7.35 \pm 0.06) \times 10^{-3}$ & $1.18 \pm 0.01$ & 2.75-3.0           & 2.25-2.5  \\
 interarm & $(8.10 \pm 0.09) \times 10^{-3}$ & $1.14 \pm 0.02$ & 2.75-3.0           & 2.25-2.5  \\
 \enddata
 \tablenotetext{a}{Uncertainties are the standard error for the means.  Calibration uncertainties are not included.}
\end{deluxetable}

\begin{deluxetable}{cccc}
\tabletypesize{\small}
\tablecolumns{4}
\tablecaption{A comparison of the PDR characteristics measured in M51 to previous studies\label{pdr_comparison}}
\tablewidth{0pt}
\tablehead{
\colhead{Paper} & \colhead{Source(s)} & \colhead{log($n$/cm$^{-3}$)} & \colhead{log$G_{0}$}}
 \startdata
 This work                   & nucleus                 & 3.75-4.0   & 3.25-3.75 \\ 
 \nodata                     & center                  & 3.0-3.25   & 2.75-3.0  \\
 \nodata                     & arms                    & 2.75-3.0   & 2.25-2.5  \\
 \nodata                     & interarm                & 2.75-3.0   & 2.25-2.5  \\
 (1) 						 & normal galaxies         & 2.0-4.0    & 2.0-4.0   \\
 (2) 						 & AGN, starbursts, normal & 2.0-4.5    & 2.0-4.5   \\
 (3) 						 & NGC~5713                & 4.2        & 2.8       \\
 (4) 						 & NGC~4214                & 3.3-3.5    & 2.9-3.0   \\
 (5) 						 & NGC~6946, NGC~1313      & 2.0-4.0    & 2.0-4.0   \\
 (6) 						 & M83, M51                & 2.0-4.25   & 2.5-5.0   \\
 (7) 						 & NGC~1097, NGC~4559      & 2.5-3.0    & 1.7-3.0   \\
 (8)			 			 & M33 (BCLMP\,302) 	   & 2.5        & 1.5 \\
 \enddata
 \tablerefs{(1) \citet{2001A&A...375..566N} (2) \citet{2001ApJ...561..766M} (3) \citet{1996A&A...315L.117L} (4) \citet{2010A&A...518L..57C} (5) \citet{2002AJ....124..751C} (6) \citet{2005A&A...441..961K} (7) \citet{2012ApJ...747...81C} (8) \citet{2011A&A...532A.152M}}
\end{deluxetable}

Finally, we compare our values for $G_{0}$ with those determined using other approaches.  First, assuming that all of our observed infrared flux has been converted from the impedent FUV flux by dust grains, we can compare the inferred values of $G_{0}$ from the PDR model to the observed $F_{\mathrm{TIR}}$ in M51.  To convert the observed integrated intensity of the TIR continuum we follow \citet{2005A&A...441..961K} and calculate $G_{0}^{obs}$ as $4\pi I_{\mathrm{TIR}}^{slab}/2(1.6 \times 10^{-3}$~erg~cm$^{-2}$~s$^{-1})$, where $I_{\mathrm{TIR}}^{slab}$ is the observed total integrated intensity reduced by a factor of two, to consider only emission from the front side of the cloud, as described above.  We find mean values for $G_{0}^{obs}$ of 120, 100, 35, and 15 for the nucleus, center, arm and interarm regions, respectively, in fairly good agreement with those determined by \citet{2005A&A...441..961K}, who found values of roughly 76, 20 and 15 for the nucleus and two locations in the spiral arms of M51.  The ratio of the observed radiation field versus that from the PDR model, $G_{0}^{obs}$/$G_{0}$, gives us a filling factor for the clouds within each beam.  We find filling factors of roughly 2--7, 10--20, 10--20, and 4--8\% for the nucleus, center, arm and interarm regions.  The lower filling factor in the interarm region is consistent with our theory that there are fewer clouds in this region than in the spiral arms.

Next, we compare the model predicted values of $G_{0}$ to the value of the FUV field determined using dust SED modelling by \citet{2012ApJ...755..165M}.  They model the SED using the model of \citet{2007ApJ...657..810D}, which parameterizes the interstellar radiation field (ISRF) with $U$, a scaling factor of the average Milky Way ISRF spectrum, $U_{\mathrm{MW}}$ from \citet{1983A&A...128..212M}. We note here that the conversion between $U$ and $G_{0}$ is $U$ = 0.88$G_{0}$ \citep{2007ApJ...663..866D}.  The ambient ISRF is represented in the dust SED model by $U_{\mathrm{min}}$. In addition, the model has a ``PDR'' component that represents a stronger radiation field due to massive stars.  This component comprises a sum of intensities with $U$ scaling factors ranging between $U_{\mathrm{min}}$ and 10$^{6}U_{\mathrm{MW}}$.  \citet{2012ApJ...755..165M} find that the typical ISRF in M51 is $\sim$5 to 10 times that of the average value in the Milky Way and their analysis covers the central region of the galaxy where we have mapped our fine-structure lines.  They also determined that the PDR component contributes at most about 2\% of the total radiation field within the same region.

Ideally, we would make a direct comparison between $G_{0}$ measured in this work and the total strength of the radiation field measured within the PDR component as measured by the dust SED model.  However, this value was not reported by \citet{2012ApJ...755..165M}.  But we do point out that the dust SED modelling suggests only 2\% of the dust content is exposed to the strong PDR component whereas we find that at least 12\% of the total dust continuum emission originates in PDRs.  This difference may arise due to the different concept of a PDR in each model.  The \citet{1999ApJ...527..795K} picture adopts a PDR as any region within the ISM where FUV photons have a significant effect on the chemistry in that region, which can include regions with weak (i.e. ambient) values of $G_{0}$ \citep{1985ApJ...291..722T}.  On the other hand, the dust SED model of \citet{2007ApJ...657..810D} assumes the PDR component is exposed to a radiation field above the ambient field, and thus may overlook some PDR regions as defined by the \citet{1999ApJ...527..795K} model.  Our models produce in some sense an average measure of $G_{0}$ over all regions within a single PACS beam (including regions with both low and high FUV radiation field strengths), and so perhaps it is not surprising that our measured values of $G_{0}$ are larger than the value of $U_{\mathrm{min}}$.

\section{Conclusions}\label{conclusions}
We present new \emph{Herschel} PACS and SPIRE observations of the grand design spiral galaxy M51 of the important fine-structure lines [C\,\textsc{ii}](158~$\mu$m), [N\,\textsc{ii}](122 \& 205~$\mu$m), [O\,\textsc{i}](63~$\mu$m), [O\,\textsc{i}](145~$\mu$m) and [O\,\textsc{iii}](88~$\mu$m).  We measure several diagnostic ratios including [C\,\textsc{ii}]/$F_{\mathrm{TIR}}$, ([C\,\textsc{ii}]+[O\,\textsc{i}]63)/$F_{\mathrm{TIR}}$, [C\,\textsc{ii}]/[O\,\textsc{i}]63, and [O\,\textsc{i}]145/[O\,\textsc{i}]63.  We find a [C\,\textsc{ii}]/$F_{\mathrm{TIR}}$ ratio of $4 \times 10^{-3}$ on average, consistent with previous results for M51, as well as other nearby galaxies in various surveys and resolved studies.  Furthermore, we see a slight deficit in this ratio in the central region of M51 when compared to the surrounding environment, and the ([C\,\textsc{ii}]+[O\,\textsc{i}]63)/$F_{\mathrm{TIR}}$ ratio suggests reduced heating efficiency in these central regions.  We also find that the [N\,\textsc{ii}]122/[N\,\textsc{ii}]205 ratio indicates that diffuse ionized gas dominates these emission lines.

We divide the disk of M51 into four regions to conduct a pixel-by-pixel analysis in each region to investigate possible variations within the properties of the interstellar gas.  We determine that the fraction of ionized gas contributing to the total observed [C\,\textsc{ii}] emission is approximately 80\% in the nucleus of the galaxy and decreases to 50\% in the arm and interarm regions.  The [O\,\textsc{i}]145/[O\,\textsc{i}]63 ratio indicates the [O\,\textsc{i}]63 line is optically thick in the inner region of M51.  We correct for both the [O\,\textsc{i}] optical depth and the ionized contribution to the [C\,\textsc{ii}] emission in our analysis.

We compare our observed line ratios in each region to the PDR model of \citet{1999ApJ...527..795K} and we find the incident FUV fluxes are approximately, $G_{0}\sim$ $10^{3.25}-10^{4.0}$, $10^{2.25}-10^{3.75}$, $10^{1.5}-10^{3.0}$ and $10^{1.5}-10^{4.0}$ for the nucleus, center, arm and interarm regions, respectively.  The density of hydrogen nuclei, $n$, is $10^{3.5}-10^{4.25}$, $10^{2.5}-10^{4.0}$, $10^{2.0}-10^{3.75}$, and $10^{2.25}-10^{4.25}$~cm$^{-3}$ for the nucleus, center, arm and interarm regions, respectively.  These derived values are similar to those previously seen in M51 on larger scales, as well as global results of numerous galaxies in surveys such as \citet{2001ApJ...561..766M}.

We show for the first time that both $G_{0}$ and $n$ decrease with increasing radius within M51.  Furthermore, we find that the arm and interarm regions, despite having different star formation rate surface densities and physical processes, show approximately the same incident FUV field and density within their molecular clouds.  Finally, our data and analysis suggest that PDRs contribute between 12\% and 50\% of the total infrared emission in M51.


\acknowledgments
T.~J.~P. would like to extend her thanks to the anonymous referee for his/her constructive comments, which have contributed to strengthening our work.  The research of C.~D.~W. is supported by the Natural Sciences and Engineering Research Council of Canada and the Canadian Space Agency.  The research of V.~L. is supported by a CEA/Marie Curie Eurotalents fellowship.  PACS has been developed by a consortium of institutes led by MPE (Germany) and including UVIE (Austria); KU Leuven, CSL, IMEC (Belgium); CEA, LAM (France); MPIA (Germany); INAF-IFSI/OAA/OAP/OAT, LENS, SISSA (Italy); IAC (Spain). This development has been supported by the funding agencies BMVIT (Austria), ESA-PRODEX (Belgium), CEA/CNES (France), DLR (Germany), ASI/INAF (Italy), and CICYT/MCYT (Spain).  SPIRE has been developed by a consortium of institutes led by Cardiff University (UK) and including Univ. Lethbridge (Canada); NAOC (China); CEA, LAM (France); IFSI, Univ. Padua (Italy); IAC (Spain); Stockholm Observatory (Sweden); Imperial College London, RAL, UCL-MSSL, UKATC, Univ. Sussex (UK); and Caltech, JPL, NHSC, Univ. Colorado (USA). This development has been supported by national funding agencies: CSA (Canada); NAOC (China); CEA, CNES, CNRS (France); ASI (Italy); MCINN (Spain); SNSB (Sweden); STFC and UKSA (UK); and NASA (USA).  HIPE is a joint development by the Herschel Science Ground Segment Consortium, consisting of ESA, the NASA Herschel Science Center, and the HIFI, PACS and SPIRE consortia.  The James Clerk Maxwell Telescope is operated by The Joint Astronomy Centre on behalf of the Science and Technology Facilities Council of the United Kingdom, the Netherlands Organisation for Scientific Research, and the National Research Council of Canada. This research has made use of the NASA/IPAC Extragalactic Database (NED) which is operated by the Jet Propulsion Laboratory, California Institute of Technology, under contract with the National Aeronautics and Space Administration.  This research made use of APLpy, an open-source plotting package for Python hosted at http://aplpy.github.com.



{\it Facility:} \facility{Herschel (PACS, SPIRE)}.



\end{document}